\documentclass[11pt]{article}
\RequirePackage[OT1]{fontenc}
\RequirePackage{amsthm,amsmath,amssymb}
\RequirePackage[colorlinks,citecolor=blue,urlcolor=blue]{hyperref}

\usepackage{graphicx,ulem}
\usepackage[utf8]{inputenc}
\usepackage{appendix}
\usepackage{enumitem}
\usepackage{dsfont}
\usepackage{comment}
\usepackage{tikz}
\usetikzlibrary{patterns}
\usetikzlibrary{arrows,intersections}
\usepackage{pgfplots}
\tikzstyle{terminal}=[circle,draw]
\pgfplotsset{width=7cm}
\newcommand{\h}[1]{\widehat{#1}}
\DeclareMathOperator*{\argmax}{arg\,max}
\DeclareMathOperator*{\argmin}{arg\,min}
\newcommand{\cqfd}{\nobreak \ifvmode \relax \else
      \ifdim\lastskip<1.5em \hskip-\lastskip
      \hskip1.5em plus0em minus0.5em \fi \nobreak
      \vrule height0.75em width0.5em depth0.25em\fi}

\makeatletter
\newcommand{\leqnomode}{\tagsleft@true}
\makeatother

\def\<{\langle}
\def\>{\rangle}

\def\e{\text{e}}

\def\Chi{\raise .3ex
\hbox{\large $\chi$}}

\def\({\Bigl (}
\def\){\Bigr )}
\newcommand{\be}{\begin{equation}}
\newcommand{\ee}{\end{equation}}

\newcommand{\f}{\frac}
\newcommand{\bea}{$$ \begin{array}{lll}}
\newcommand{\eea}{\end{array} $$}
\newcommand{\bi}{\begin{itemize}}
\newcommand{\ei}{\end{itemize}}

\newtheorem{definition}{Definition}
\newtheorem{remark}{Remark}

\DeclareMathOperator{\R}{{\mathbb R}}

\begin{document}

%
\title{Testing for high frequency features in a noisy signal}
%
%
%
%
%
%
%

\author{ Mathieu Mezache\footnote{{ Sorbonne Universités, Inria, Universit\'{e} Paris-Diderot, CNRS,  Laboratoire Jacques-Louis Lions, F-75005 Paris, France,mathieu.mezache@inria.fr}}
\and Marc Hoffmann\footnote{Universit\'e Paris-Dauphine PSL, CEREMADE, Place du Mar\'echal de Lattre de Tassigny, F-75016 Paris, hoffmann@ceremade.dauphine.fr}
\and Human Rezaei\footnote{INRA, UR892, Virologie Immunologie Moléculaires, 78350 Jouy-en-Josas, France, human.rezaei@inra.fr  } 
\and Marie Doumic\footnote{{Sorbonne Universités, INRIA, Universit\'{e} Paris-Diderot, CNRS,  Laboratoire Jacques-Louis Lions, F-75005 Paris, France, marie.doumic@inria.fr - Wolfgang Pauli Institute, c/o university of Vienna, Austria}}  
}
\date{}
\maketitle

\begin{abstract}
Given nonstationary data, one generally wants to extract the trend from the noise by smoothing or filtering. However, it is often important to delineate a third intermediate category, that we call {\it high frequency (HF) features}: this is the case in our motivating example, which consists in experimental measurements of the time-dynamics of depolymerising protein fibrils average size. One may intuitively visualise HF features as the presence of fast, possibly nonstationary and transient oscillations, distinct from a slowly-varying trend envelope. The aim of this article is to propose an empirical definition of HF features and construct estimators and statistical tests for their presence accordingly, when the data consists of a noisy nonstationary $1$-dimensional signal. We propose a parametric characterization in the Fourier domain of the HF features by defining a maximal amplitude and distance to low frequencies of significant energy. We introduce a data-driven procedure to estimate these parameters, and compute a p-value proxy based on a statistical test for the presence of HF features.
 The test is first conducted on simulated signals where the ratio amplitude of the HF features to the level of the noise is controlled. The test detects HF features even when the level of noise is five times larger than the amplitude of the oscillations. In a second part, the test is  conducted on experimental data from Prion disease experiments and it confirms the presence of HF features in these signals with significant confidence.
\end{abstract}





\section*{Introduction}

\subsection*{Motivation}

\begin{figure}[htb!]
\begin{center}
\scalebox{0.7}{\includegraphics[width=\textwidth]{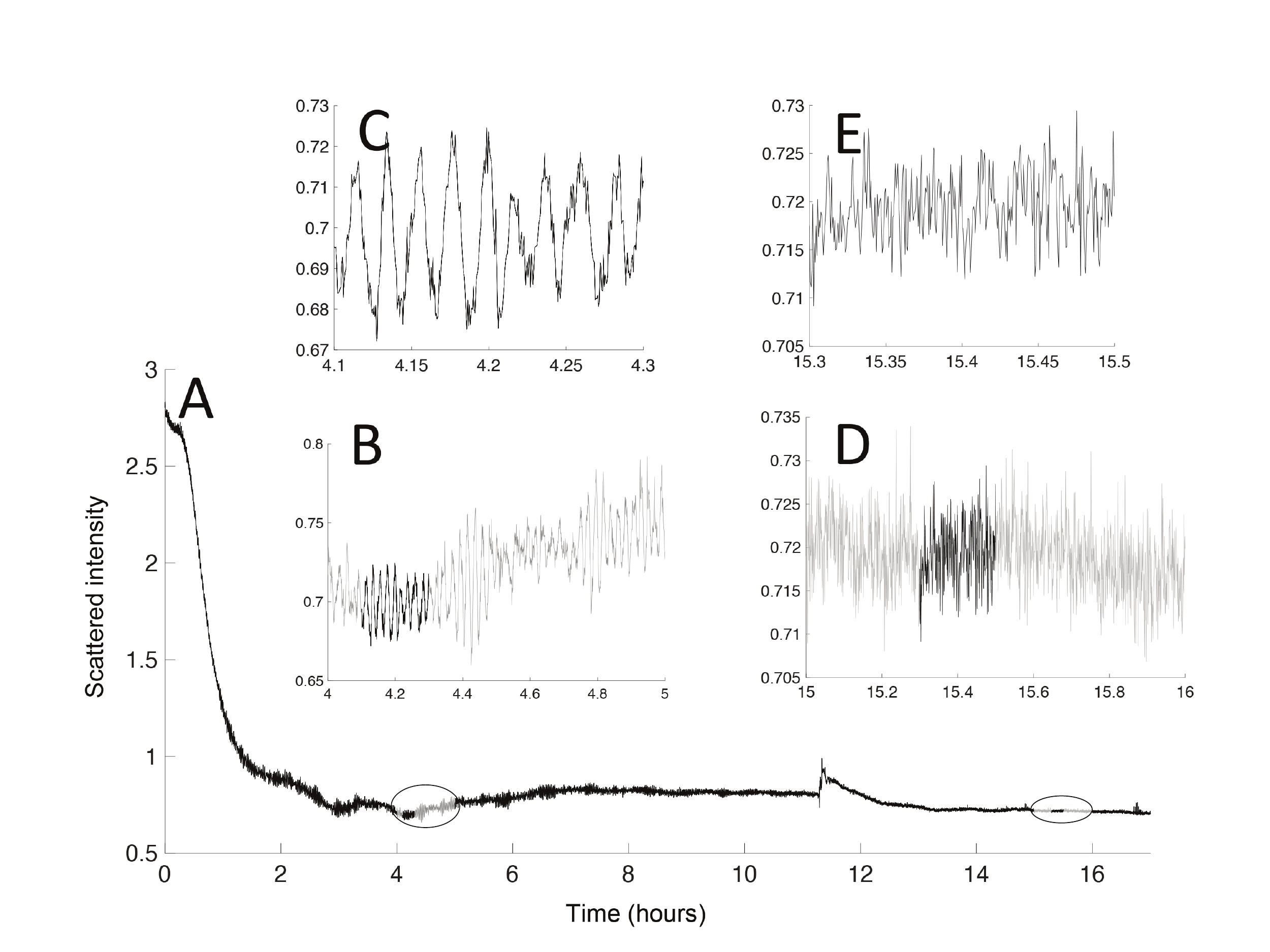}}
\caption{\label{fig:DFMR}
{\bf  Human PrP amyloid fibrils (Hu fibrils) depolymerisation monitored by Static Light Scattering} (see Appendix for details). A:  The overall view of the $0.35\mu M$ Hu-fibrils depolymerisation at $55^0 C$. B-E correspond to a zoom-in on different time-segments of the depolymerisation curve A. As  shown in B, from time 4h to time 5h  oscillations have been observed when  for time segment corresponding to time 15.3 to 15.5h only noise has been detected (D). (Figure taken from~\cite{doumic:hal-01863748}) }
\end{center}
\end{figure}

In a one-dimensional nonstationary signal, it is often of key importance to delineate not only the trend of the signal from its noise components - which is an extensively studied subject, see {\it e.g.} \cite{Pinsker, LMS, donoho1994ideal, donoho1998minimax}  or the classical textbooks \cite{N, HKPT} and the references therein - but also to delineate high-frequency features from low frequency characteristics. As an example, and original motivation for our study, fast oscillations have been visually observed in experimental measurements of the infectious agent in Prion diseases, see Figure~\ref{fig:DFMR}~B and C; they are however not easily quantified or distinguished from zones where only noise is present, as in Figure~\ref{fig:DFMR}, D and E.

There are two main difficulties in order to infer such oscillations and evaluate their significance: first, they are mixed up with noise and second, they need to be separated from the trend of the signal. It is therefore of major interest to rely on a systematic procedure that estimates quantitatively high frequency (HF) features - amplitude, frequency - in real signals and thus delineates these features of the signal from both pure noise and trend. \\

Studies on spectral analysis of a signal are usually based on stationary or weakly stationary models, see {\it e.g.} Chapter 4 in \cite{shumway2017time}, or~\cite{koopmans1995spectral}. 
The Singular Spectrum Analysis (SSA) introduced in~\cite{broomhead1989time}  allows one to visualize qualitative dynamics from noisy experimental data. The SSA is based on the decomposition of a time series or signal into several additive components interpreted as trend components, oscillatory components, and noise components. It is then widely used to identify intermittent or modulated oscillations in time series, see e.g. ~\cite{vautard1992singular, paluvs2008detecting,golyandina2001analysis}. A statistical test of hypothesis to discriminate between potential oscillations and noise has been introduced in ~\cite{allen1996monte} and~\cite{palus1998detecting}. This test is called the Monte Carlo SSA and has been applied almost exclusively to meteorological data. Since SSA transforms the original data in a complex way, no theoretical result has yet been proved on the Monte Carlo SSA. Prior knowledge on the signal (such as the trend or assumptions on the noise) are also needed in order to calibrate the procedure and improve the result of the statistical test. The Monte Carlo SSA is by construction a non-parametric procedure and the oscillations detected by this test are not characterized quantitatively but qualitatively.\\ 

However,  the experimental signals studied in this paper have local oscillations and exhibit a nonstationary behaviour. Hence, classical methods seem inappropriate for our class of signals. In this paper, we propose a new method, relying on nonlinear analysis of the Fourier transform of the signal that enables one to infer a data-driven parametric characterization of HF features, based on their amplitude and frequency detection without any {\it a priori} knowledge of their shape or location. The main idea is to make use of the lowest frequency features, which are the ones with highest amplitude in the signals studied, and to detect the peak with highest amplitude situated at some distance of them in the frequency axis. This method is detailed in Section~\ref{sec:Char}. We  introduce a  statistical test to discriminate HF features from noise in Section~\ref{sec:process_pval}, apply our methodology to a simulated example in Section~\ref{sec:num}, and then to the experimental measurements of depolymerising PrP protein fibrils displayed in Figure~\ref{fig:RD_trend} in Section~\ref{sec:PrP}.  \\

\subsection*{Model and assumptions}

For some (large) $n \geq 1$, we have measurements $y_i^n$ of a noisy signal localized around $i/n$, so that $i$ is a location parameter and $n$ a frequency parameter. We may idealise our data via a representation of the form
\begin{equation} \label{eq: data gen process}
y_i^n=x_i^n+\sigma \xi_i^n,\qquad i= 0,\hdots,n-1,
\end{equation}
\noindent
where 
$(x_i^n)_{0 \leq i \leq n-1}$ is the true (unknown) signal of interest and the $\xi_i$ are independent and identically distributed noise measurements, that we assume here to be standard Gaussian. The quantity $\sigma >0$ is a (fixed) noise level. In this nonparametric regression setting, we aim at detecting from the data $(y_i^n)_{0 \leq i \leq n-1}$ whether $(x_i^n)_{0 \leq i \leq n-1}$ exhibits {\it high-frequency features} (HF features) such as oscillations, a term that still needs to be defined properly.  Since we do not know in advance whether such high-frequency features are present and where they are located, we need to investigate the shape of $(x_i^n)_{0 \leq i \leq n-1}$, which requires some smoothing in order to get rid of the noise $(\xi_i^n)_{0 \leq i \leq n-1}$. However, any smoothing procedure tends to wipe out high-frequencies in the data, which is adversarial to our goal, therefore the choice of the smoothing method is crucial. 

\subsection*{Results and organisation of the paper}

The statistical test to differentiate HF features from noise in a signal is data-driven and is based on the study of the projection of the signal in the Fourier domain. We propose in Section~\ref{sec:Char} a parametric characterization of the HF features of a signal. This characterization also provides with an algorithmic procedure for the computation of the HF features, implemented in the Python language at 
\begin{center}
\texttt{https://github.com/mmezache/HFFTest} 
\end{center}
see Appendix ~\ref{app:LibPyth}.
The construction of the statistical test of hypothesis and the computation of a p-value proxy is described in Section~\ref{sec:process_pval}.
The numerical examples are performed in Section~\ref{sec:num} on simulated signals. They are constructed around parameters which control their trend, their 
high frequency feature component and their noise. We vary the ratio of the amplitude of the HF features over the noise level (i.e. its standard deviation), which sheds light on the robustness of the procedure: the transient oscillations are detected by the procedure even if the noise level is significantly high.
The procedure is then applied to static light scattering (SLS) experiments of $PrP^{Sc}$ fibrils, in Section~\ref{sec:PrP}. They are characterised by their singular slow-varying components (non-monotonous trend) and their fast-varying components (isolated discontinuous jumps, high frequency features, noise). We compute the HF features parameters of SLS signal experiments for different initial concentration of $PrP^{Sc}$. We conclude that these signals have significant HF features, i.e. the signals display transient oscillations coming from biochemical reactions and not from the experimental noise.

\section{Characterisation of high frequency features}\label{sec:Char}

\subsection*{Notation, graphical definition and guidelines of the procedure}
%
Let $(x_i^n)_{0 \leq i \leq n-1}$ be  a real-valued discrete signal of length $n$, and let us denote  $\big(\vartheta_{n,k}\big)_{0 \leq k \leq n-1}$ its discrete Fourier transform  (DFT), classically defined by
\begin{equation} \label{eq: FFT}
\mathsf{DFT}_n\big[(x_i^n)_{0 \leq i \leq n-1}\big] = \Big(\sum_{i = 0}^{n-1} x_i^n\e^{-j2\pi k i/n}\Big)_{0 \leq k \leq n-1} = \big(\vartheta_{n,k}\big)_{0 \leq k \leq n-1}.
\end{equation}

\begin{figure}[!htb]
\centering
\includegraphics[height=0.5\linewidth]{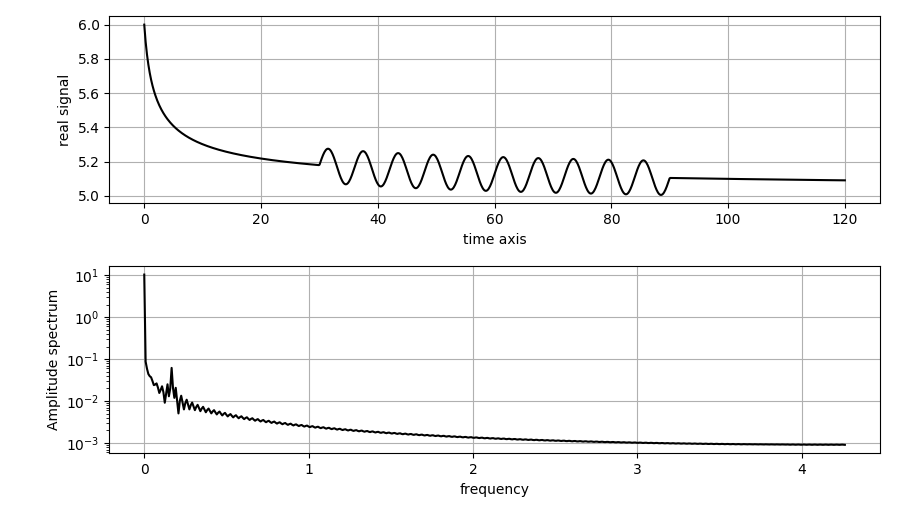}
\caption{{\footnotesize {\bf Graph of a test signal with HF features and its single-sided amplitude spectrum. } {\bf Top:} Plot of $z_i=f(0.12i)$ for $ i=0,\hdots,1024 $ where  $f(x)=\f{1}{\sqrt{x+1}}+0.5\sin\left(\f{2\pi x}{6}\right)\mathds{1}_{[30,90]}(x)+5$. {\bf Bottom:} Plot of the amplitude spectrum of $(z_i)_{0\leq i\leq 1024}$ (logarithmic scale for the y-axis). \label{fig:DFT_lowFreqOsc}}}
\end{figure}

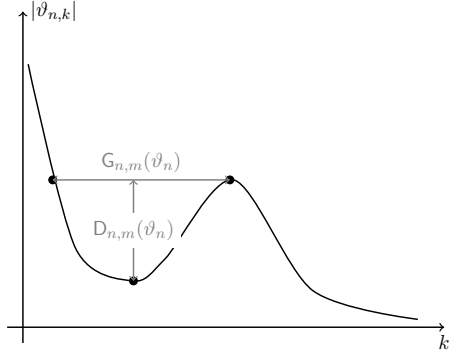
\begin{figure}[!htb]
        \centering
\scalebox{0.7}{%
\begin{tikzpicture}[
    thick,
    dot/.style = {
      draw,
      fill = white,
      circle,
      inner sep = 0pt,
      minimum size = 4pt
    }
  ]
  \coordinate (O) at (0,0);
  \coordinate (Min) at (2.1,0.88);
  \coordinate (Max) at (2.1,2.8);
  \draw[->] (-0.3,0) -- (8,0) coordinate[label = {below:$k$}] (xmax);
  \draw[->] (0,-0.3) -- (0,6) coordinate[label = {right:$|\vartheta_{n,k}|$}] (ymax);
  \path[name path=x] (0,2.8) -- (8,2.8);
  \path[name path=y] plot[smooth] coordinates {(0.1,5) (1,1.5) (2.7,1.3) (4,2.8) (5.5,0.7) (7.5,0.15)};
  \scope[name intersections = {of = x and y, name = i}]
    \fill[gray!20] (i-1) -- (i-2 |- i-1) -- (i-2) -- cycle;
    \draw[black] plot[smooth] coordinates {(0.1,5) (1,1.5) (2.1,0.88) (2.7,1.3) (4,2.8) (5.5,0.7) (7.5,0.15)};
    \filldraw (i-1) circle (2pt);
	\filldraw (i-2) circle (2pt);
	\filldraw (Min) circle (2pt);
    \draw[gray, <->] (Max) -- node[fill = white]
      {$\mathsf D_{n,m}(\vartheta_n)$}  (Min);	
	\draw[gray, <->] (i-1)--node[above] {$\mathsf G_{n,m}(\vartheta_n)$} (i-2);    
  \endscope
\end{tikzpicture}
}
\caption{{\footnotesize  {\bf Idealised scheme of the parametrization of the HF features of a signal in the Fourier Domain.} The parameter $\mathsf G_{n,m}(\vartheta_n)$ is the location parameter in the frequency scale which corresponds to the distance of the HF features from the low-frequency components of the signal. The parameter $\mathsf D_{n,m}(\vartheta_n)$ is the intensity parameter which corresponds to the relative amplitude of the HF features. }
\label{fig:scheme_par_osc}}
\end{figure}

Recommended reference on applied Fourier analysis is the textbooks of \cite{STEIN, ARFKEN} or \cite{shumway2017time} in a time series context. Our experimental signals have a specific low frequency trend combined with HF features  that shall persist beyond denoising. The presence of a trend implies that  there are 
large Fourier coefficients $\vartheta_{n,k}$ on the scale corresponding to the lowest frequency information, whereas HF features 
can be characterised by {\it relatively} large coefficients in  higher frequencies
As displayed by the test signal in Figure~\ref{fig:DFT_lowFreqOsc}, a typical signal displaying oscillations would thus consist, in the frequency domain, of large coefficients in the lowest frequencies,  a decay to a local minimum,  then one or more peaks in higher frequencies, and a final decay or plateau corresponding to the noise amplitude in the highest frequencies. In order to discreminate between significant and nonsignificant peaks (straightforward to detect visually in Figure~\ref{fig:DFT_lowFreqOsc}), we also need a preliminary smoothing step: we then obtain an amplitude spectrum that looks similar to the scheme displayed in Figure~\ref{fig:scheme_par_osc}.

Given a discrete signal $(x_i^n)_{0 \leq i \leq n-1}$, with DFT $\vartheta_n = (\vartheta_{n,k})_{0 \leq k \leq n-1}$ defined by Equation~\eqref{eq: FFT}, we are thus led to characterise a HF feature by two nonnegative parameters: 
\begin{enumerate}
\item A location parameter $\mathsf G_{n,m}(\vartheta_n)$ defining a distance between the highest peak and  the range of lowest frequencies, 
\item an amplitude parameter $\mathsf{D}_{n,m}(\vartheta_n)$, measuring the relative amplitude of the peak and the first local minimum. 
\end{enumerate}
We have displayed their graphical definition in Figure~\ref{fig:scheme_par_osc}. The parameters $\mathsf G_{n,m}(\vartheta_n)$ and $\mathsf D_{n,m}(\vartheta_n)$  depend on a smoothing parameter $m$ and represent two distances: $\mathsf G_{n,m}(\vartheta_n)$ is a distance on the frequency axis and $\mathsf D_{n,m}(\vartheta)$ on the amplitude axis. For each signal, the parametric characterization is unique: it describes the peak with the highest distance between its amplitude and the minimum amplitude of the Fourier coefficients of lower frequencies (with $\mathsf D_{n,m}(\vartheta_n)$). The parameter $\mathsf G_{n,m}(\vartheta_n)$ gives the distance in frequency indices between the peak and the components in the low frequencies with the same intensity.

The algorithmic definition of $\mathsf G_{n,m}(\vartheta_n)$ and $\mathsf D_{n,m}(\vartheta_n)$  is provided in the following subsection, Definition~\ref{def: characteristics} below. As already said, it depends not only on $n$ but also on a pre-regularization parameter $m.$ As classical in many other smoothing situations (see {\it e.g.} \cite{donoho1994ideal, donoho1998minimax} or the textbook \cite{TSYB}), the choice of this parameter is crucial, since it has to be large enough  to smooth the lowest frequency range into a monotonously decaying curve, and small enough to avoid altering too much the HF features. A data-driven procedure to define $m$ is explained in Section~\ref{sec:num}, Equation~\eqref{def:Smooth_par}.

\subsection*{Algorithmic procedure}

 We have explained so far the guidelines of the procedure and provided a graphical definition of the HF parameters. Let us now detail the algorithmic procedure. In order to be fully rigorous (for instance to be certain of selecting a unique peak in case of several peaks of the same amplitude), this algorithm contains technical details which may seem cumbersome at first sight; it is possible to omit it at first reading. The  first step of the procedure is a preliminary smoothing step; the second step consists of the detection and localization of all peaks of amplitude which follow the first local minimum; finally, the peak of highest amplitude is selected. 

\subsubsection*{First step: Pre-processing the signal}

Replacing $x_{i}^n$ by $x_{i}^n+C$ for some arbitrary constant $C$, with no loss of generality, we may (and will) assume that 
\begin{equation} \label{eq: init}
|\vartheta_{n,0}| > \max_{0 \leq k \leq n-1}|\vartheta_{n,k}|.
\end{equation}
\noindent
Condition \eqref{eq: init} is in force from now on. 
We transform $\vartheta_{n} =  (\vartheta_{n,k})_{0 \leq k \leq n-1}$ into a non-decreasing sequence $\mu_{n}^{(m)} = (\mu_{n,k}^{(m)})_{m \leq k \leq n-m-1}$ that depends on a smoothing parameter $m$ (with  $0 \leq m \leq \f{n-1}{2}$) defined as follows:
\begin{equation}\label{def:orderstat}\mu_{n,m}^{(m)} = \min_k \vartheta_{n,k}^{(m)} \leq \mu_{n,m+1}^{(m)} \leq \ldots \leq \mu_{n,k}^{(m)} \leq \mu_{n,n-m}^{(m)} = \max_k \vartheta_{n,k}^{(m)}
\end{equation}
where

\begin{equation}
\vartheta_{n,k}^{(m)} = \Big(\tfrac{1}{2m+1}\sum_{l = k-m}^{k+m}|\vartheta_{n,l}|^2\Big)^{1/2},\;\;m \leq k \leq n-m-1.\label{def:CoefFour_regul}
\end{equation}
 \noindent
In other words, the sequence $\mu_{n}^{(m)}$ is the order statistics of a $2m$-regularised version of $\vartheta_{n}$.

\begin{figure}[!htb]
\centering
\includegraphics[height=0.6\linewidth]{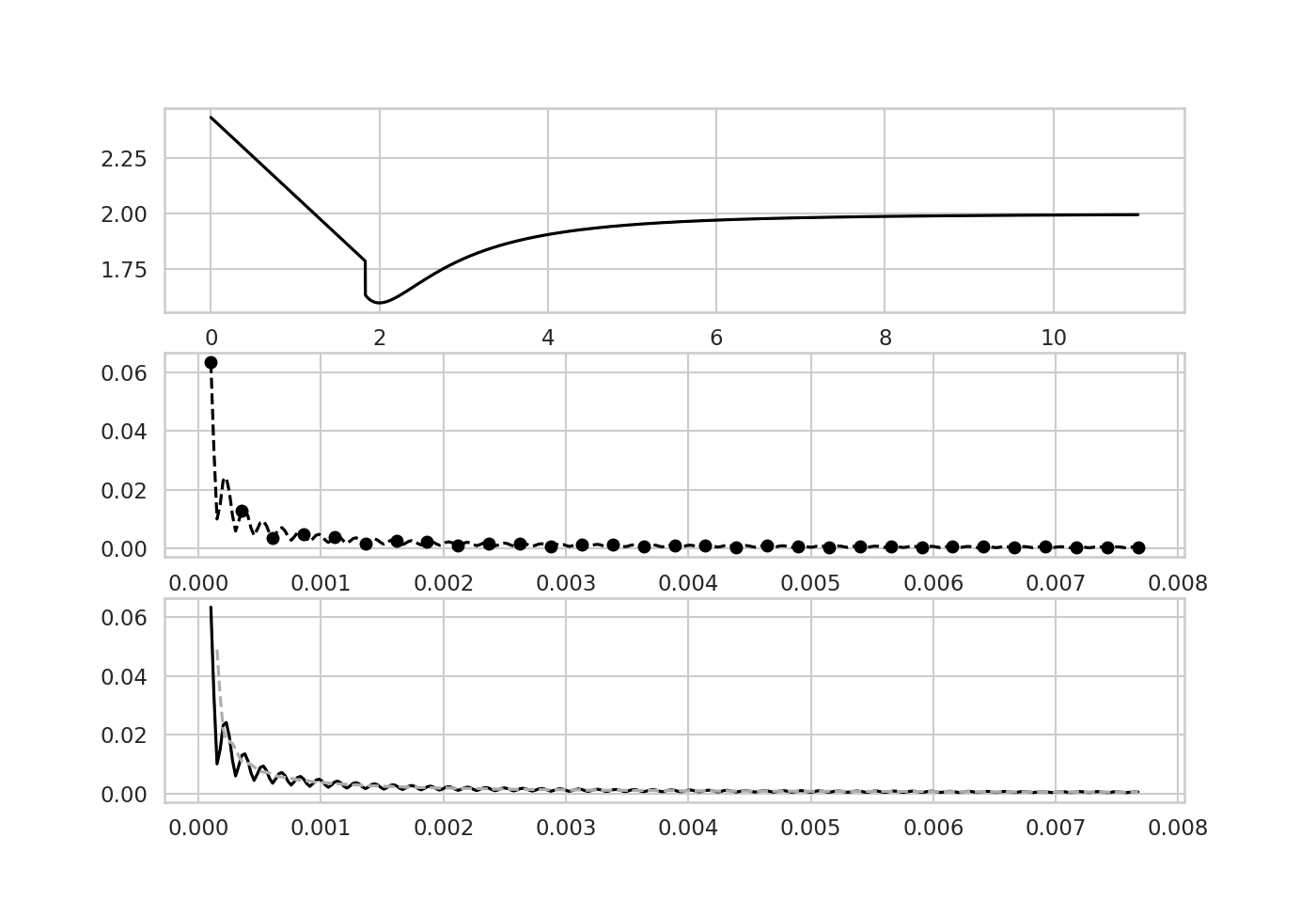}
\caption{{\footnotesize {\bf Graph of a test signal with a jump and a change of monotonicity and its single-sided amplitude spectrum. } {\bf Top:} Graph of the signal with a decreasing, increasing and stationary part. {\bf Middle:} Zoom on the low frequency of the amplitude spectrum for $n=10000$ samples of the signal. The dot markers emphasize one over ten samples of the signal. {\bf Bottom:} Plot of the amplitude spectrum of the test signal (plain line). The dash line corresponds to the plot of $(\vartheta_{n,k}^{(3)})_{3\leq k \leq n-4}$ defined by \eqref{def:CoefFour_regul} with $m=3$.} \label{fig:DFT_lowFreqjump}}
\end{figure}

\begin{remark} The smoothing parameter $m$ is needed as soon as the signal observed displays singularities e.g. a jump discontinuity or a fast transition of monotonicity of the trend. These phenomena are approximated by the harmonic sequence $(\e^{j2\pi k\cdot}, k\in \mathbb Z)$, and when projected in the Fourier domain, the amplitude spectrum displays a serie of spikes (cf Figure~\ref{fig:DFT_lowFreqjump}). These phenomena are related to Gibbs phenomenon (\cite{zygmund2002trigonometric}, Chapter 2) and give rise to spikes in the Fourier domain which can be falsely interpreted as HF features. The regularization with an adequate choice of the parameter $m$ solves this issue (cf Figure~\ref{fig:DFT_lowFreqjump} and Section~\ref{sec:num}). Moreover, any regularization of the signal hinders the detection of the HF features since it smoothes the signal. The regularization method defined in \eqref{def:CoefFour_regul}, is certainly not the only possible one, but it has the advantage of giving the right trade-off between smoothing and not altering too much the original signal in order to detect the HF features of the experimental data studied. In Section \ref{sec:num}, Equation~\eqref{def:Smooth_par}, we also detail a procedure for a data-driven choice of $m$. \label{rmk:par_m} 
\end{remark}

\begin{remark}
The regularization of order $2m$ transforms the sequence $\vartheta_{n}$ of $n$ terms into a sequence of $n-2m$ terms in order to avoid boundary effects. We label the indices of the series from $m$ to $n-m-1$ in so that the parameter $k$ in $\vartheta_{n,k}^{(m)}$ is reminiscent of a frequency parameter and we formally have $\vartheta_{n,k}^{(0)} = |\vartheta_{n,k}|$.
\end{remark}

\subsubsection*{Second Step: Detection and Localization of significant features in the Fourier domain; choice of the peak of highest amplitude}

\begin{figure}[!htb]
        \centering
\scalebox{0.7}{%
\begin{tikzpicture}[
    thick,
    dot/.style = {
      draw,
      fill = white,
      circle,
      inner sep = 0pt,
      minimum size = 4pt
    }
  ]
  \coordinate (O) at (0,0);
  \coordinate (Min) at (2.1,0.88);
  \coordinate (Max) at (1.6,2.8);
  \coordinate[label={left:$\mu_l$}] (muj) at (0,2.8);
  \coordinate[label={left:$\mu_i$}] (mui) at (0,0.88);
  \draw[->] (-0.3,0) -- (8,0) coordinate[label = {below:$k$}] (xmax);
  \draw[->] (0,-0.3) -- (0,6) coordinate[label = {right:$|\vartheta_{n,k}|$}] (ymax);
  \path[name path=x] (0,2.8) -- (8,2.8);
  \path[name path=y] plot[smooth] coordinates {(0.1,5) (1,1.5) (2.7,1.3) (4,2.8) (5.5,0.7) (7.5,0.15)};
  \scope[name intersections = {of = x and y, name = i}]
    \fill[gray!20] (i-1) -- (i-2 |- i-1) -- (i-2) -- cycle;
    \draw      (0,2.8) -- (8,2.8);
    \draw      (0,0.88)-- (8,0.88);
    \draw[black] plot[smooth] coordinates {(0.1,5) (1,1.5) (2.1,0.88) (2.7,1.3) (4,2.8) (5.5,0.7) (7.5,0.15)};
    \filldraw (i-1) circle (2pt);
	\filldraw (i-2) circle (2pt);
	\filldraw (Min) circle (2pt);
	\draw[black,dashed](Min) -- (2.1,0)coordinate[label = {below:$a(\mu_i) $}] (del);
	\draw[black,dashed](i-2) -- (4,0)coordinate[label = {below:$b(\mu_i)=b(\mu_l) $}] (delu);
	\draw[dashed](i-1) -- (0.75,0)coordinate[label = {below:$a(\mu_l) $}] (del);
  \endscope
\end{tikzpicture}
}
\caption{{\footnotesize 
Illustration of the procedure where $\mu_l$ is the largest level of amplitude for which $b(\mu_l)$ is large enough to quit the low amplitude range (for $\mu_p>\mu_l,$ we have $a(\mu_p)=b(\mu_p)$); 
we have $i<l,$ $\mu_i= \min\limits_{m \leq k \leq \mathsf b (\mu_l)}\vartheta_{n,k}^{(m)},$ $\; a(\mu_i)> a(\mu_l) \; \text{and} \; b(\mu_i)=b(\mu_l)$.}}
\label{fig:fourier}
\end{figure}
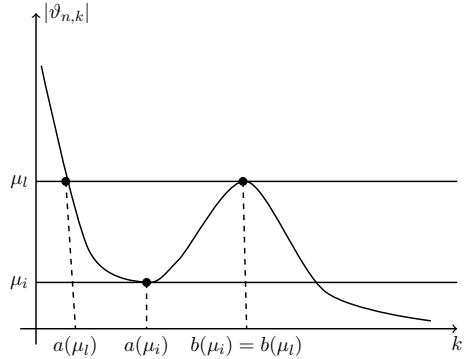

Define, for $x \geq 0$
\begin{equation} \label{eq: def a}
\mathsf a(x) = \mathsf a_{n}^{(m)}(x) = \min \big\{k\;|\; m \leq k \leq n-m-1,\;\vartheta_{n,k}^{(m)} \leq x\big\}
\end{equation}
\noindent
and
\begin{equation} \label{eq: def b}
\mathsf b(x) = \mathsf b_n^{(m)}(x) =\max\bigg\{ \argmax \big\{\vartheta_{n,k}^{(m)}\; | \;\mathsf a(x) \leq k \leq n-m-1\big\}\bigg \}.
\end{equation}
\begin{remark}
The algorithm (see the summary below) starts to compute $\mathsf a$ and $\mathsf b$ at $x=\mu_{n,n-m-1}^{(m)}$ and then makes it decrease to $x=\mu_{n,m}^{(m)}.$ For large values of $x,$ we have $\mathsf a=\mathsf b,$ situated in the decreasing part of the signal in the lowest frequency range. Then it reaches amplitudes which are attained at more than one place, and we get $\mathsf a (x) < \mathsf b(x)$. The index $\mathsf a(x)$ is the minimal frequency at which searching for HF features starts, getting rid of the potentially high energy levels arising from the low frequency part of the signal. The index $\mathsf b(x)$ is a maximal frequency for which the energy level $x$ is reached in the search zone $\{\mathsf a(x), \mathsf a(x)+1,\ldots, n-m-1\}$. See also Figure~\ref{fig:fourier}.
\end{remark}

Define the sets
\begin{equation}\label{def:setA}
\mathcal A_{n}^{(m)} = \big\{\mu_{n,k}^{(m)}\; | \; \mu_{n,k}^{(m)} = \vartheta^{(m)}_{n,\mathsf b(\mu_{n,k}^{(m)} )},\;m \leq k \leq n-m-1\big\}
\end{equation}
and
\begin{equation}\label{def:setS}
\mathcal S_{n}^{(m)} = \big\{\mu_{n,k}^{(m)} \in \mathcal A_n^{(m)}\; | \; \mathsf b(\mu_{n,k}^{(m)}) > \mathsf a(\mu_{n,k}^{(m)}),\;m \leq k \leq n-m-1\big\}.
\end{equation}

\begin{remark}
The set $\mathcal A_n^{(m)}$ represents potential candidates for maximum energy levels of a HF feature, while $\mathcal S_n^{(m)}$ represents the set of amplitudes of the spikes of $\vartheta_{n} $.
\end{remark}


To define the HF features, we now select in the set $\mathcal S_n^{(m)} $ the feature with maximum relative amplitude. Let us define
\begin{equation} \label{eq: def d}
\mathsf d(x) = \mathsf d_n^{(m)}(x) = x-\min_{m \leq k \leq \mathsf b_n^{(m)} (x)}\vartheta_{n,k}^{(m)}
\end{equation}
\noindent
and we obtain a maximum intensity of HF feature as
\begin{equation}\label{def:iota}
\iota_{n}^{(m)}(\vartheta_{n}) = \left\{\begin{array}{l}
\max \bigg \{\argmax_{x \in \mathcal S_n^{(m)}}\mathsf d_n^{(m)}(x)\bigg\}\quad 
\text{if }\mathcal S_n^{(m)}\text{ is non empty,}
\\0\qquad \text{otherwise.}
\end{array}\right.
\end{equation}
if $\mathcal S_n^{(m)}$ is non empty and $\iota_{n}^{(m)}(\vartheta_{n})=0$ otherwise. Moreover, if the set $\argmax_{x \in \mathcal S}\mathsf d_n^{(m)}(x) $ is not reduced to a singleton, taking its maximum ensures us to obtain a unique element for $\iota_{n}^{(m)}(\vartheta_{n}) $ i.e. the feature of maximum relative amplitude and maximum intensity. We are ready to give a quantitative definition of a HF feature:

\begin{definition} \label{def: characteristics}To any discrete signal  $\vartheta_n = (\vartheta_{n,k})_{0 \leq i \leq n-1}$ given in the Fourier domain, we associate a high-frequency feature (HF feature) $\big(\mathsf G_{n,m}(\vartheta_n), \mathsf D_{n,m}(\vartheta_n)\big)$ at discretisation level $n \geq 1$ and smoothing level $m \leq \tfrac{n-1}{2}$ as follows:
$$\mathsf G_{n,m}(\vartheta_n) =  \mathsf b_n^{(m)}\big(\iota_n^{(m)}(\vartheta_n)\big)  -  \mathsf a_n^{(m)}\big(\iota_n^{(m)}(\vartheta_n)\big)$$
and
$$\mathsf D_{n,m}(\vartheta_n) = \mathsf d_n^{(m)}\big(\iota_n^{(m)}(\vartheta_n)\big),$$
where $\mathsf a_n^{(m)},$ $\mathsf b_n^{(m)}$,   $\mathsf d_n^{(m)}$ and $\iota_n^{(m)}$ are defined in \eqref{eq: def a}, \eqref{eq: def b},~\eqref{eq: def d} and~\eqref{def:iota} respectively. 

\end{definition}

\subsection*{Summary of the algorithm}
We assume here that we are given a signal $(y_ i^n)$ for $0\leq i\leq n-1$ and a smoothing parameter $m.$
\begin{enumerate}
\item Define $(\vartheta_{n,k})_{0\leq k\leq n-1}$ by~\eqref{eq: FFT}.
\item Define $\vartheta_{n,k}^{(m)},$  $m\leq k\leq n-m-1,$ by~\eqref{def:CoefFour_regul}.
\item Define $\mu_{n,k}^{(m)},$  $m\leq k\leq n-m-1,$ by~\eqref{def:orderstat}.
\item For  $k=n-m,\cdots,m,$ define $\mathsf a(\mu_{n,k}^{(m)})$ by~\eqref{eq: def a} and $\mathsf b(\mu_{n,k}^{(m)})$ by~\eqref{eq: def a}.
\item Define the sets $\mathcal A_{n}^{(m)}$ by~\eqref{def:setA} and $\mathcal S_{n}^{(m)}$ by~\eqref{def:setS}.
\item Define $\iota_n^{(m)}$ by~\eqref{def:iota}.
\item Define $\mathsf G_{n,m}(\vartheta_n)$ and $\mathsf D_{n,m}(\vartheta_n) $ by Definition~\ref{def: characteristics}.
\end{enumerate}

\section{Testing for HF features}\label{sec:process_pval}
For any given signal $(y_ i^n)$, we have  defined  two HF parameters $\mathsf G_{n,m}(\vartheta_n)$ and $\mathsf D_{n,m}(\vartheta_n)$ in a unique way. The question to answer is then: are these two parameters high enough to really characterise significant HF features, or are they insignificant compared to the noise level of the signal considered? The difficulty is to know whether the amplitude of the peak selected and its distance to the  range of the lowest frequencies are significant or not, so that we may conclude that the signal does display or does not display HF features. The aim of this section is to  define a statistical test in order to give a quantitative answer to this question.

We continue with the statistical setting introduced in Equation \eqref{eq: data gen process}:
we observe
\begin{equation*} 
y_i^n=x_i^n+\sigma \xi_i^n,\quad i= 0,\hdots,n-1,
\end{equation*}
\noindent
where $(x_i^n)_{0 \leq i \leq n-1}$ is the signal of interest and $(\sigma\xi_i^n)_{\leq i\leq n}$ are independent centred Gaussian random variables with noise variance $\sigma^2$, for some (large) $n \geq 1$, interpreted as a maximal discretisation resolution level or equivalently a maximal frequency of observation. Applying the discrete Fourier transform $\mathsf{DFT}_n$ on both sides of \eqref{eq: data gen process}, we equivalently observe
$$
\widehat \vartheta_{n,k} = \vartheta_{n,k} + \sigma \widetilde \xi_{k,n}, \quad k= 0,\hdots,n-1,
$$
where the $\sigma\widetilde \xi_{k,n}$ are independent centred Gaussian random variables with variance $\sigma^2$ as well, thanks to the fact that $\mathsf{DFT}_n$ is an orthogonal linear mapping.
From data $(y_i^n)_{0 \leq i \leq n-1}$ or rather $(\widehat \vartheta_{n,k})_{0 \leq k \leq n-1}$, we wish to construct a statistically significant test of the absence of HF feature as the null, against a set of local alternatives where some HF features are present.\\

We first formulate the hypothesis of the test based on the HF parameters according to Definition \ref{def: characteristics}. We then construct the statistical test  and define the test statistics.  We define empirically the risk region under the null thanks to a Monte Carlo method. This method is also assimilated to bootstrap hypothesis testing, see e.g. \cite{efron1992bootstrap}, \cite{fisher1990bootstrap}, and it is motivated by its simplicity and the fact that the theoretical distribution is  unknown. Moreover it is a straightforward way to generate the empirical distribution of the test statistics under the null hypothesis. Finally, we compute the test statistics of the observations, its p-value proxy and obtain a decision rule of the statistical test.

\subsection{Construction of a statistical test}

Thanks to the characterisation of HF features 
via $\big(\mathsf D_{n,m}(\vartheta_n),\mathsf G_{n,m}(\vartheta_n)\big)$ given in Definition~\ref{def: characteristics},
we test the null
$$\mathcal H_{n,m,\nu,c}^{0}:\;\;\mathsf G_{n,m}(\vartheta_n)<\nu,\quad  \mathsf D_{n,m}(\vartheta_n)< c,$$
against the local alternatives
$$\mathcal H_{n,m,\nu, c}^1:\;\mathsf G_{n,m}(\vartheta_n) \geq \nu \;\;\text{and}\;\;\mathsf D_{n,m}(\vartheta_n) \geq c, $$
where $\nu>0,\; c>0$ are thresholds to determine significant HF features. The null hypothesis $\mathcal H^{0}$ is that there is no significant HF feature in the signal tested. On the contrary, the hypothesis $\mathcal H^{1}$ implies that the signal has significant HF feature. For the test to be powerful, the main problem is to define $(\nu, c)$: for too small values any signal shall reject $\mathcal{H}^0$ whereas for large values, any signal shall accept $\mathcal{H}^0$. We obtain a test statistics for $\big(\mathsf G_{n,m}(\vartheta_n),\mathsf D_{n,m}(\vartheta_n)\big)$ by setting
$$\widehat{\mathsf G}_{n,m} = \mathsf G_{n,m}(\widehat \vartheta_n) =  \mathsf b_n^{(m)}\big(\iota_n^{(m)}(\widehat \vartheta_{n})\big)  -  \mathsf a_n^{(m)}\big(\iota_n^{(m)}(\widehat \vartheta_{n})\big)$$
and
$$\widehat{\mathsf D}_{n,m} = \mathsf D_{n,m}(\widehat \vartheta_n) =  \mathsf d_n^{(m)}\big(\iota_n^{(m)}(\widehat \vartheta_{n})\big).$$
In order to compute the p-value proxy of the test, we design a Monte Carlo procedure simulating a proxy of the data $(y_i)_{0\leq i\leq n-1}$ under the null $\mathcal{H}^0$. Using the proxy, we define a reject region of our test for a risk level $\alpha$ and the p-value proxy of the data $(y_i)_{0\leq i\leq n-1}$.

\subsubsection*{Rejection zone at risk level $\alpha$}

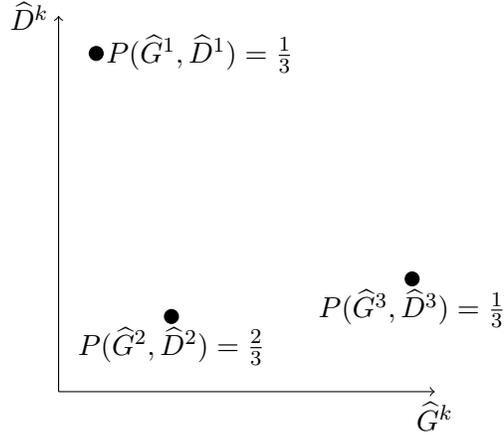
\begin{figure}[!htb]
        \centering
\begin{tikzpicture}
\draw [->] (0,0)--(0,5);
\draw [->] (0,0)--(5,0);
\path (5,0) node [anchor=north] {$\widehat{G}^k$};
\path (0,5) node [anchor=east] {$\widehat{D}^k$};
\path (0.5,4.5) coordinate (gd1);
\path (1.5,1) coordinate (gd2);
\path (4.7,1.5) coordinate (gd3);
\draw (gd1) -- (gd1) node [anchor=west] {$P(\widehat{G}^1,\widehat{D}^1)=\tfrac{1}{3}$};
\draw (gd2) -- (gd2) node [anchor=north] {$P(\widehat{G}^2,\widehat{D}^2)=\tfrac{2}{3}$};
\draw (gd3) -- (gd3) node [anchor=north] {$P(\widehat{G}^3,\widehat{D}^3)=\tfrac{1}{3}$};
\fill (gd1) circle (1mm);
\fill (gd2) circle (1mm);
\fill (gd3) circle (1mm);
\end{tikzpicture}
        \caption{{\bf Cloud of points $\big(\widehat{\mathsf G}^k, \widehat{\mathsf D}^k\big)=\left(\mathsf G_{n,m}\left(\mathsf{DFT}[y_{\lambda,n}^{(0),k}]\right),\mathsf D_{n,m}\left(\mathsf{DFT}[y_{\lambda,n}^{(0),k}]\right)\right)$ for $k=1,2,3$.} }
        \label{Scheme:set_threshold}         
\end{figure}

We first simulate $N$ times $y^{(0)}_{\lambda, n}$ defined in \eqref{eq: simul null} below, which is a simulated proxy of the data $(y_i^n)_{0 \leq i \leq n-1}$ with HF features removed from the signal $(x_i^n)_{0 \leq i \leq n-1}$.
Repeating independently $N$ times the procedure, we obtain a Monte Carlo sequence
$$y^{(0),k}_{\lambda, n}\; k=1,\hdots, N.$$  
In a second step, we denote by $E_N^0$ the cloud of points representing the HF features parameters of these simulated signals (with HF features removed but with Gaussian noise):
\begin{equation}
E_N^0=\left\{\left(\mathsf G_{n,m}\left(\mathsf{DFT}[y_{\lambda,n}^{(0),k}]\right),\mathsf D_{n,m}\left(\mathsf{DFT}[y_{\lambda,n}^{(0),k}]\right)\right) \; | \; k=1,\hdots,N \right\}.
\label{def:Set_threshold}
\end{equation}
\noindent
We define the function $P: \R_+^2 \rightarrow F\subset [0;1]$:
\begin{equation}
P(g,d)=N^{-1}\sum_{k = 1}^N {\bf 1}_{\left\{\mathsf G_{n,m}\left(\mathsf{DFT}[y_{\lambda,n}^{(0),k}]\right) \geq g,\; \mathsf D_{n,m}\left(\mathsf{DFT}[y_{\lambda,n}^{(0),k}]\right) \geq d\right\}}.\label{def:func_P}
\end{equation}
Hence $ P(g,d)$ is an empirical probability that corresponds to the proportion of points in $E_N^0$ located in the North-East quarter of the plane centered on $(g,d)$ (cf Figure~\ref{Scheme:set_threshold}). In order to reduce the computation cost, we only consider the restriction of $P$ to the set $E_N^0$. Thus 
if $E_N^0$ contains $N$ disjoint points then the minimal bound on $P(E_N^0) $ is $\tfrac{1}{N}$.
For a risk level $\alpha \in P(E_N^0) $, the rejection zone of our test is defined as
\begin{equation}
\mathcal R_{m,n}\big(\kappa_1^\alpha,\kappa_2^\alpha\big) = \big\{(y_i)_{1\leq i\leq n}\text{ defined by~\eqref{eq: data gen process} s.t. }\widehat{\mathsf G}_{n,m} \geq \kappa_1^\alpha,\;\;\widehat{\mathsf D}_{n,m} \geq \kappa_2^\alpha\big\}
\label{eq: def risque 1}
\end{equation}
\noindent
where $ (\widehat{\mathsf G}_{n,m},\widehat{\mathsf D}_{n,m})$ is the test statistics and $\big(\kappa_1^\alpha, \kappa_2^\alpha\big) \in E_N^0$ are such that
\begin{equation}
\label{set_threshold}
P( \kappa_1^\alpha,  \kappa_2^\alpha)= \alpha.
\end{equation}
\begin{remark}
The risk level $\alpha$ is imposed by the Monte Carlo sequence, $\alpha \in P(E_N^0)  \subset [\tfrac{1}{N};1]$. For example, Figure~\ref{Scheme:set_threshold} represents an arbitrary set $E_N^0$ for $N=3$. We note that we need $\tfrac{1}{3}\leq \alpha \leq 1$ in order to obtain candidates $\kappa_1^\alpha, \kappa_2^\alpha$. For $\alpha < \tfrac{1}{3}$ no candidate can be obtained by this procedure and its associated reject region is not defined. Moreover there can be multiple reject regions defined for the same risk level $\alpha$ (in the example whe have two reject regions for $\alpha=\tfrac{1}{3}$).
\end{remark}

The main idea behind the computation of the couples $\left(\mathsf G_{n,m}\left(\mathsf{DFT}[y_{\lambda,n}^{(0),k}]\right),\mathsf D_{n,m}\left(\mathsf{DFT}[y_{\lambda,n}^{(0),k}]\right)\right)$ is to generate random outcomes under the null $\mathcal H^{(0)}$ that enable us to compute an empirical risk level. The couples correspond to the relative amplitude and the frequency gap for a non-oscillating signal with noise. We also get reject region(s) of level $\alpha$ thanks to the threshold(s) $\big( \kappa_1^\alpha, \kappa_2^\alpha\big)$. We do not need uniqueness of the reject region in order to define and compute the p-value proxy, see below.

\subsubsection*{Definition of the $\text{p-value}$ proxy}
We define, for  observations $(y_i^n)_{0\leq i \leq n-1}$:
\begin{equation}
\text{p-value proxy}\big((y_i^n)_{0 \leq i \leq n-1}\big) = \min \big\{ \alpha\in P(E_N^0) \;|\;\widehat{\mathsf G}_{n,m}\geq \kappa_1^\alpha,\quad\widehat{\mathsf D}_{n,m}\geq \kappa_2^\alpha\big\}.
\label{def:p-val}
\end{equation}
\noindent
An equivalent definition of the \text{p-value proxy} of the observations $(y_i^n)_{0\leq i \leq n-1}$ is given by
$$\text{p-value proxy}\big((y_i^n)_{0 \leq i \leq n-1}\big) = \inf \big\{\alpha \in P(E_N^0)  | \;(y_i^n)_{0 \leq i \leq n-1} \in \mathcal R_{m,n}\big(\kappa_1^\alpha,\kappa_2^\alpha\big)\big\}.$$
\begin{figure}[!htb]
        \centering
\begin{tikzpicture}
\draw [->] (0,0)--(0,5);
\draw [->] (0,0)--(5,0);
\path (5,0) node [anchor=north] {$\widehat{G}_{n,m}^k$};
\path (0,5) node [anchor=east] {$\widehat{D}_{n,m}^k$};
\path (0.5,4.5) coordinate (gd1);
\path (1.5,1) coordinate (gd2);
\path (4.7,1.5) coordinate (gd3);
\path (3,3) coordinate (pval);
\draw (gd1) -- (gd1) node [anchor=west] {$\left(\kappa_1^{1/3},\kappa_2^{1/3}\right)$};
\draw (gd2) -- (gd2) node [anchor=north] {$\left(\kappa_1^{2/3},\kappa_2^{2/3}\right)$};
\draw (pval) -- (pval) node [anchor=north] {$\big(\widehat{\mathsf G}_{n, m}, \widehat{\mathsf D}_{n, m}\big)$};
\draw (gd3) -- (gd3) node [anchor=north] {$\left(\kappa_1^{1/3},\kappa_2^{1/3}\right)$};
\fill (gd1) circle (1mm);
\fill (gd2) circle (1mm);
\fill (gd3) circle (1mm);
\fill [color=gray] (pval) circle (1mm);
\end{tikzpicture}
        \caption{{\bf Cloud of points $\big(\widehat{\mathsf G}_{n, m}^k, \widehat{\mathsf D}_{n, m}^k\big)$
        for $k=1,2,3$ (black dots) and HF feature parameters of the observations (the test statistics) $\big(\widehat{\mathsf G}_{n, m}, \widehat{\mathsf D}_{n, m}\big)$
         (grey dot).}}
        \label{Scheme: CloudOfPoints 2}         
\end{figure}
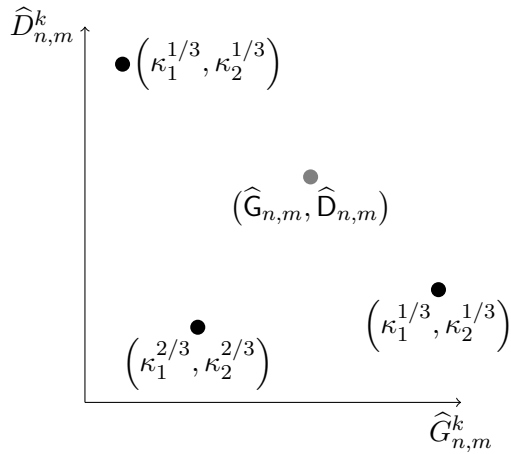
\begin{remark}
The computation of the $\text{p-value proxy}$ is illustrated schematically in Figure~\ref{Scheme: CloudOfPoints 2}. We observe the point cloud formed by $\big(\widehat{\mathsf G}_{n, m}^k, \widehat{\mathsf D}_{n, m}^k\big)$
 (the black dots) for $k=1,2,3$. On the vertical axis we have the relative amplitude and on the horizontal axis we have the gap in frequency between the oscillations and the trend in the Fourier domain. The grey dot illustrates the HF features parameters subject to the test. We use the 
  $\big(\widehat{\mathsf G}_{n, m}^k, \widehat{\mathsf D}_{n, m}^k\big)$ 
as our grid to compute the p-value proxy. Using \eqref{set_threshold}, we compute the level $\alpha$ and obtain consequently the $\left(\kappa_1^\alpha,\kappa_2^\alpha\right)$ for each element of the grid. The p-value proxy is $\tfrac{2}{3}$ in this example.
\end{remark}

In order to conclude, we first need to define $\alpha^{*}$ corresponding to the probability threshold below which the null hypothesis $\mathcal H^{(0)}$ is be rejected. Using \eqref{def:Set_threshold} and \eqref{set_threshold}, we define the significance level $\alpha^*$ as $$\alpha^* = \min\lbrace P(E_N^0)\rbrace.$$
In particular, $ \alpha^* =\tfrac{1}{N}$ will stand true in the numerical examples and in the analysis of the experimental signals (cf. Section \ref{sec:num}). Hence, we reject the null hypothesis in favor of the alternative hypothesis if and only if the p-value proxy defined in \eqref{def:p-val} is less or equal to (equal to, in Section \ref{sec:num}) $\alpha^*$. The interpretation of this decision rule is schematized in Figure~\ref{Scheme: CloudOfPoints 2}. The decision to reject $\mathcal H^{(0)}$ is equivalent to the fact that the couple formed by the test statistics $\left( \widehat{\mathsf G}_{n, m}, \widehat{\mathsf D}_{n, m}\right) $ is located in the north-east quarter of the plane centered on a boundary point of the cloud of points defined by \eqref{def:Set_threshold}. In the example in Figure \ref{Scheme: CloudOfPoints 2}, the decision is to not reject the null $\mathcal H^{(0)} $ since the test statistics couple is "located in" the cloud of points.
\\

The $\text{p-value}$ proxy gives a confidence index for non-rejecting the null. This index is meaningful provided the test has a good power, {\it i.e.} if the probability of making a type II error is small.
Hence the $\text{p-value proxy}$ of $\big((y_i^n)_{0 \leq i \leq n-1}\big)$ is our measure of confidence in non-rejection of the null $\mathcal H^0$. The main difficulty however lies in solving \eqref{eq: def risque 1} since $\vartheta_n$ remains unknown under the null and that there are no reason that $\widehat{\mathsf G}_{n,m}$ or $\widehat{\mathsf D}_{n,m}$ are pivotal statistics under the null. We describe below a numerical procedure based on Monte Carlo simulation that estimates $y_{\lambda,n}^{(0)} $ a proxy of the data with HF features removed but with noise.

\subsection{A Monte Carlo procedure for the simulation of the null}\label{sec:MC process}

In order to evaluate \eqref{eq: def risque 1} and \eqref{def:p-val},  we first build a low-frequency estimator $\widehat x_{\lambda, n}^{(0)}$ from the data $(y_i^n)_{0 \leq i \leq n-1}$ that removes the potential HF features. The estimator depends on a regularisation parameter $\lambda$. We next define
\begin{equation} \label{eq: simul null}
y^{(0)}_{\lambda, i,n}  = \widehat x_{\lambda, i, n}^{(0)}+\widehat \sigma_n \epsilon_i^n,\;\;i =0,\ldots, n-1,
\end{equation}
\noindent
where the $\epsilon_i^n$ are independent centred Gaussian random variables that we simulate and $\widehat \sigma_n$ is an estimator of the standard deviation of the noise. 
The simulated signal $(y^{(0)}_{\lambda, i,n})_{0 \leq i \leq n-1}$ obtained by estimating a proxy of $f$ with HF features removed with additional simulated noise serves as a proxy of the data $(y_i^n)_{0 \leq i \leq n-1}$ under the null $\mathcal H^0$. 

\subsubsection*{Numerical computation of $\widehat f_{\lambda, n}^{(0)}$}
Trend estimation or filtering for mimicking a signal with HF features removed has many applications and hence it has been extensively studied.


In the following, the trend is considered as the underlying slowly varying component of the signal and we choose the $\ell_1$-trend filtering method described in~\cite{kim2009ell_1} to estimate it (see Appendix~\ref{app:trend} for details on this method and the reasons of our choice). The estimator of $\widehat x_{\lambda, n}^{(0)}$ as a $n$-dimensional vector is then the solution of the following optimisation problem:
\begin{equation}
\widehat x_{\lambda, n}^{(0)} \in \argmin\limits_{x\in\mathbb{R}^n}\frac{1}{2}\sum\limits_{i=1}^{n-1} (y_i^n-x_i^n)^2+\lambda\sum\limits_{i=1}^{n-2}|x_{i-1}^n-2x_i^n+x_{i+1}^n|,
\label{def:l1Trend_est}
\end{equation}
\noindent
where $\lambda\geq 0$ is a regularisation parameter which controls the trade-off between the smoothness of $\widehat x_{\lambda, n}^{(0)}$ and the residual $\sum_{i = 0}^{n-1}\big(\widehat x_{\lambda, n,i}^{(0)}-y_i^n\big)^2$. We note that the second term $ \sum\limits_{i=1}^{n-2}|x_{i-1}^n-2x_i^n+x_{i+1}^n|$ is the $\ell^1 $-norm of the second order variations of the sequence $(x^n)$ (i.e. the discretization of the corresponding $L^1$-norm of the second derivative of a function). 
Since there is no optimal criterium to choose $\lambda$, the choice of the parameter is qualitative and motivated empirically (see Section~\ref{sec:num} and Appendix~\ref{app:trend}).
 
\subsubsection*{Numerical estimation of the noise level $\widehat \sigma_n$}
The estimator of the standard deviation of the noise is the second ingredient needed in order to compute $\widehat f_{\lambda, n}^{(0)}$ in \eqref{eq: simul null}. The methods to estimate the level of noise are closely linked to the methods of signal denoising and have been extensively studied. The method chosen here  is the median absolute deviation and the denoised signal is obtained thanks to the wavelet shrinkage methods, see {\it e.g.}~\cite{donoho1994threshold,donoho1998minimax,donoho1995wavelet,donoho1994ideal}. We detail the method in Appendix~\ref{app:noise}.

\section{A simulation example for a proof of concept}\label{sec:num}

\subsubsection*{Pre-processing: a data-driven choice of $m$}


We first address the delicate issue of choosing the smoothing parameter $m$. Define a sequence $(m_i)_{1 \leq i \leq K}$ such that
$$1 = m_1 < m_2 < \ldots < m_K \leq \tfrac{n-1}{2}.$$
We can take for instance $m_i= i$ for $i=1,\hdots,K$. Note that $K\in \lbrace 1,\hdots,\tfrac{n-1}{2} \rbrace$ is the parameter defining the length of the finite sequence $(m_i)_{1 \leq i \leq K}$. This parameter can be fixed by the user in order to reduce the number of iterations of the procedure to compute the HF features. However, a standard choice of $K$ to obtain a data-driven procedure is $K=\tfrac{n-1}{2},$  since averaging the signal over more than half of the sample size is obviously meaningless. A good rule of thumb is $K=n^{\f{1}{2}}$, since it reduces the number of computations and remains relevant compared to the range of the signal.
Let  
$$i^\star \in \argmax_{1 \leq i \leq K}\big|\widehat{\mathsf G}_{n,m_i}-\widehat{\mathsf G}_{n,m_{i-1}}\big|$$
and we choose $m$ as $\widehat m$ defined by
\begin{equation}
\widehat m=\left\{
\begin{array}{lll}
m_{i^\star} & \text{if} & \widehat{\mathsf G}_{n,m_{i^\star}} > \widehat{\mathsf G}_{n,m_{i^\star-1}} \\
m_{i^\star-1} & \text{otherwise.} &
\end{array}
\right.\label{def:Smooth_par}
\end{equation}
As previously stated in Remark~\ref{rmk:par_m}, the empirical signals observed are non-monotonous, contain singularities and transient oscillations. Their amplitude spectra display a series of spikes in the low-frequencies and in the mid or high frequencies. Hence without a pre-processing step, the HF feature parameters (Definition \eqref{def: characteristics}) characterize the low frequencies features (i.e. the trend represented in the amplitude spectrum by spikes in the low frequencies, see Figure~\ref{fig:DFT_lowFreqjump}).\\

In order to solve this problem, we regularize the Fourier coefficients as defined in~\eqref{def:CoefFour_regul}. The sequence $(m_k)_{1\leq k\leq K}$ gradually smoothes the Fourier amplitude spectrum: the spikes in the low frequencies merge together whereas the isolated spikes in the mid or high frequencies (corresponding to transient oscillations) slightly decrease in amplitude but remain significant. The data-driven choice of $m$ is well adapted to regularize the empirical signals since it chooses the parameter $\widehat m$ from the sequence $(m_k)_{1\leq k\leq K}$ which maximizes the difference between the localisation parameters $\widehat{{\mathsf G}}$ for two consecutive smoothing parameters. Thus the spikes located in a close frequency range have been smoothed and the remaining spikes of significant amplitude for the regularization parameter $\widehat{m}$ are isolated in the Fourier amplitude spectrum.

\subsubsection*{Defining a test signal}

\begin{figure}[!htb]
\centering
\includegraphics[height=0.5\linewidth]{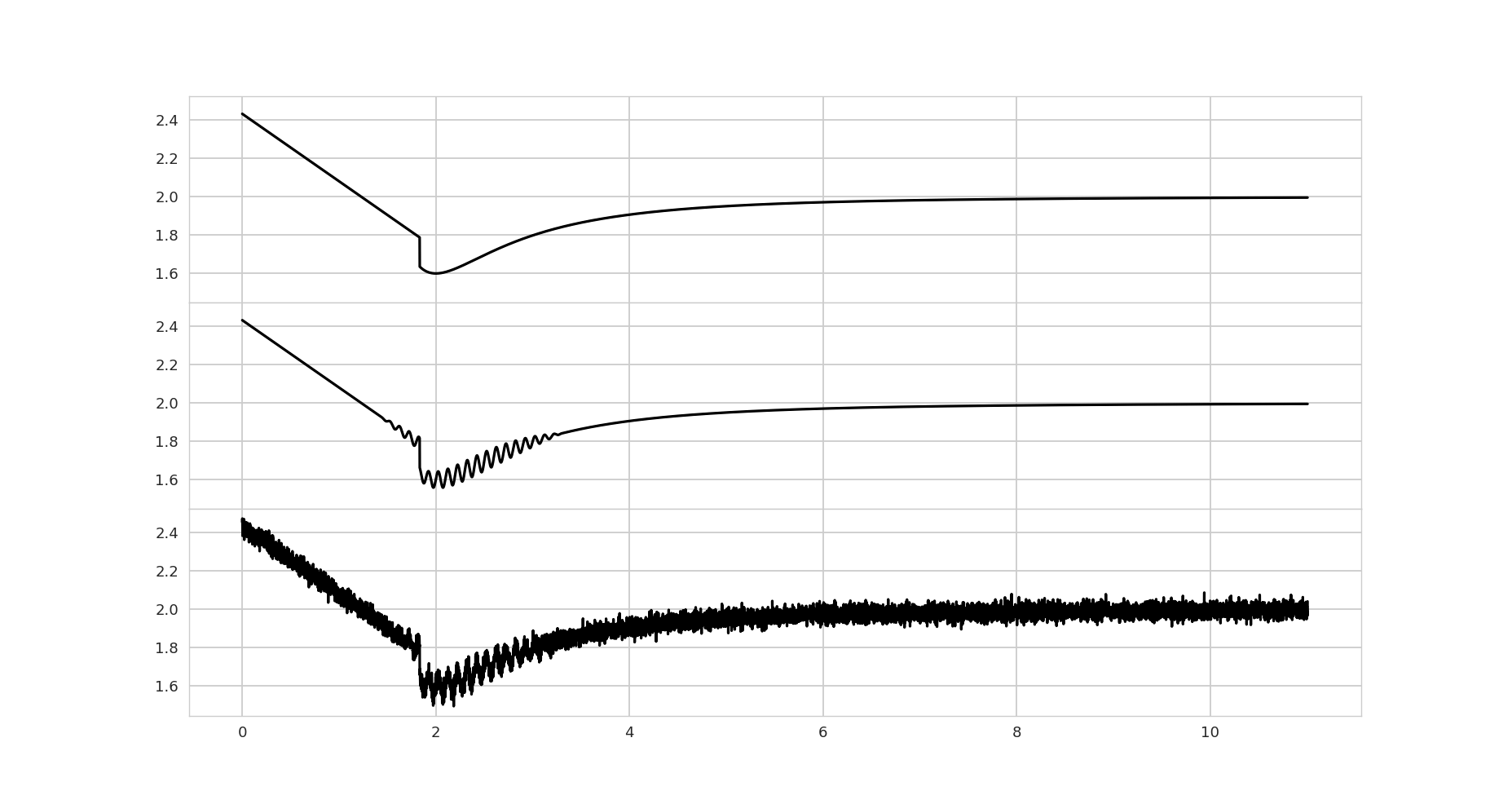}
\caption{{\footnotesize {\bf Simulation of the test signal defined by \eqref{Def:SignalSC}.}The x-axis is the time in hours. (Up) Plot of $(T_i)_{0\leq i \leq 10^5}$ with parameters (see Appendix~\ref{app:test} for details) $c_1=0.4,\; c_2=c_3=c_4=2,\; \frac{p}{2}=q=3, j_0=1700,\; j_1=3400$. (Middle) Plot of $(T_i+ O_i)_{0\leq i \leq 10^5}$ with the same parameters and $c_a=0.05,\; c_f=10$. (Down) Plot of $(S_i)_{0\leq i \leq 10^5}$ with the same parameters and $\sigma=0.025$.} \label{fig:SC_Signal}}
\end{figure}

To study numerically the validity of the procedure and the statistical test, we first compute a simulated signal where all the parameters are known. To do so, we superimpose three signals: one for the general trend of the curve, one for the HF features, and one for the noise. The signal obtained is the vector $(S_i)_{0\leq i\leq n-1}$:
 \begin{equation}
S_i=T_i+O_i+\sigma \xi_i,
\label{Def:SignalSC}
\end{equation}
\noindent
where $\sigma>0$ is the parameter corresponding to the level of noise and $\xi_i$ are  realisations of independent and identically normally distributed  random variables. Moreover $(T_i)_{0\leq i\leq n-1}$ corresponds to the trend and $(O_i)_{0\leq i\leq n-1} $ to the HF features (cf Figure~\ref{fig:SC_Signal}) 
We postpone to Appendix~\ref{app:test} the full definition of the signal, which involves three characteristic parameters:
 \begin{itemize}
\item $\sigma$ the standard deviation of the normal distributed noise,
\item $c_a$ the parameter corresponding to the amplitude of the oscillations,
\item $c_f$ the parameter corresponding to the frequency of the oscillations (since the time scale is in hours, $c_f/3600 $ is expressed in Hz).
\end{itemize}

\paragraph{Numerical computations and robustness of the procedure.} We want to understand the robustness of the numerical procedure when the frequencies and the amplitudes of the oscillations are fixed but the level of noise varies. Other said, for which parameters of the oscillations and for which level of noise does the test return that the signal oscillates (or not)? In order to answer this question, we propose the following sensitivity analysis.
The smoothing parameter $\widehat{m}$ is chosen thanks to the data-driven procedure described previously \eqref{def:Smooth_par}. 
The relevant output of our model is the p-value proxy of the signals computed thanks to the numerical procedure. A natural way to study the sensitivity of the p-value proxy to the parameters is to fix all parameters but one and observe the effect on the p-value proxys obtained. In this example the varying parameter is the level of noise $\sigma \in \left\lbrace \frac{1}{10}c_a,\; \frac{1}{2}c_a,\; c_a,\; 2c_a,\; 10c_a\right\rbrace.$
\begin{figure}
\centering
\begin{tabular}{c|c c c c c}

\\
$\sigma$ & $\frac{1}{10}c_a $ & $ \frac{1}{2}c_a$ & $ c_a$ & $2 c_a $ & $10 c_a $\\
\\ \hline
\\ 
$\widehat{m}$ & $6$ & $6$ & $6$ & $6 $ & $24$\\
 $\widehat{\mathsf G}_{n,\widehat m}$ (Hz)&$2.069\e{-3}$ &$2.044\e{-3}$ & $2.145\e{-3}$ &$1.943\e{-3}$ &$8.437\e{-2}$ \\ 
$\widehat{\mathsf D}_{n,\widehat m}$ &$1.73\e{-4}$  &$1.807\e{-4}$ & $1.807\e{-4}$&$1.844\e{-4}$ &$2.768\e{-3}$ \\
 p-value proxy& $5\e{-5}$& $5\e{-5}$&$5\e{-5}$ &$5\e{-5}$ & $4.023\e{-1}$\\
\end{tabular}
\caption{{\bf Table of estimators and p-value proxys of the simulated signals}. The simulation of the null is performed with the real trend of the signals.}
\label{tab:SC}
\end{figure}
\begin{figure}[!htb]
\centering
\includegraphics[height=0.5\linewidth]{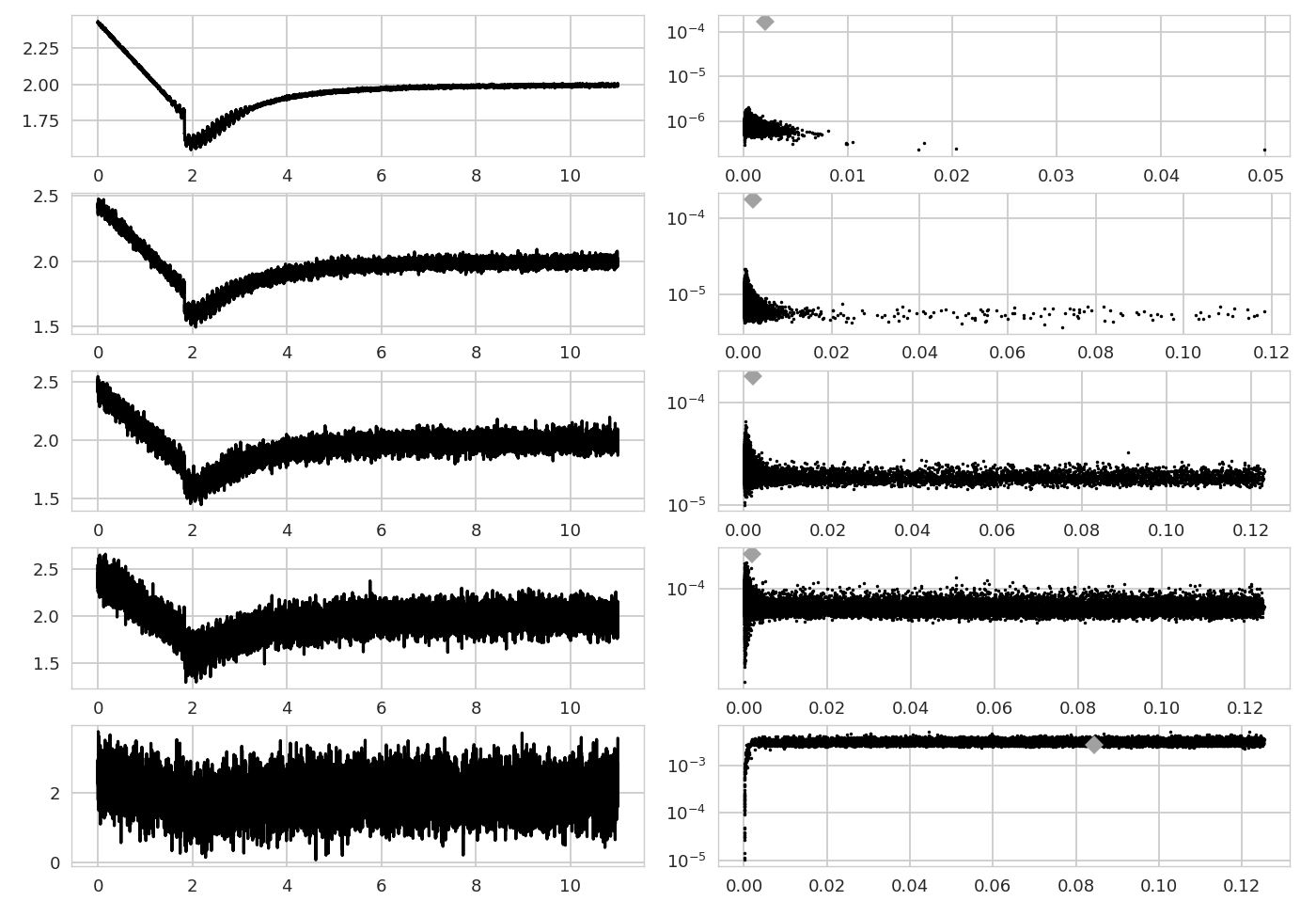}
\caption{{\footnotesize {\bf Numerical results of the procedure on the test when the simulation of the null is performed with the real trend. }(Left column) Plot of $(S_i)_{1\leq i \leq 10^5}$ \eqref{Def:SignalSC} with the parameters $c_1=0.4,\; c_2=c_3=c_4=2,\; \frac{p}{2}=q=3, t_0=1.43,\; t_1=3.30,$ $c_a=0.05,\; c_f=10$ and $\sigma\in \left\lbrace \frac{1}{10}c_a,\; \frac{1}{2}c_a,\; c_a,\; 2c_a,\; 10c_a\right\rbrace$ from top to bottom. The x-axis is the time in hours. (Right column) The black dots are the cloud of points of the simulation of the null, for $N=20000$. The grey diamond corresponds to the HF features parameters of the corresponding signal on the left column. The x-axis is the localization parameters $\widehat{G}_{n,\widehat{m}} $ and the y-axis is the relative amplitude $\widehat{D}_{n,\widehat{m}} $.
 \label{fig:SC_test}}}
\end{figure}

\subsubsection*{First test} 
Since we are working with a constructed test signal, we obtain $(\widehat{\mathsf G}_{n,\widehat m}^k, \widehat{\mathsf D}_{n,\widehat m}^k)$ in Figure~\ref{fig:SC_test} by applying the procedure of detection of the HF feature parameters setting $c_a=c_f=0$ (it corresponds to $S_i =T_i+\sigma \xi_i$ in \eqref{Def:SignalSC}). Thus the simulation of the null in Section~\ref{sec:MC process} is performed using the real trend of the signal in \eqref{eq: simul null}. Then the signals tested (Figure~\ref{fig:SC_test}) are constructed signal with parameters $c_a=0.05$, $ c_f=10$ and $ \sigma \in \lbrace \frac{1}{10}c_a,\; \frac{1}{2}c_a,\; c_a,\; 2c_a,\; 10c_a \rbrace$ in \eqref{Def:SignalSC}. The results of the detection of HF features  and the statistical test are in Table~\ref{tab:SC}. We note that for standard deviations of the noise between a tenth and the double of the amplitude of the oscillations, the p-value proxy of the test is equal to $5\e-5$. Hence, we are inclined to reject the hypothesis $\mathcal{H}^0$ which corresponds to the hypothesis that the signal displays no oscillations. Moreover we note that the signals with  standard deviations of the noise between $\frac{1}{10}c_a$ and $2c_a$ have almost the same HF feature parameters where $(\widehat{\mathsf G}_{n,\widehat m}, \widehat{\mathsf D}_{n,\widehat m})\approx (2\e-3,1.8\e-4)$. In contrast, for the signal with the standard deviation of the noise of $10c_a$, the p-value proxy is equal to $0.4$, hence we are inclined to accept that the signal has not significant enough HF feature.

\subsubsection*{Second test} 
The second step is to test the procedure on the same signals but using the trend estimate given by \eqref{def:l1Trend_est} and the noise estimation procedure described in the first step of Section~\ref{sec:MC process}. 
As displayed in Figure~\ref{fig:SC_trend} of Appendix~\ref{app:trend}, although the trend estimation quality decreases as the standard deviation of the noise increases, it remains qualitatively correct to estimate the trend of a signal displaying jumps or spikes.  Therefore we compute the procedure to obtain the HF features parameters for the simulated signals using \eqref{Def:SignalSC} with standard deviation level $\sigma \in \left\lbrace \frac{1}{10}c_a,\; \frac{1}{2}c_a,\; c_a,\; 2c_a,\; 10c_a\right\rbrace. $ The p-value proxys are computed using the $\ell_1$-trend estimators in order to obtain the couples $(\widehat{\mathsf G}_{n,\widehat m}^k, \widehat{\mathsf D}_{n,\widehat m}^k)$ where $k=1,\hdots,20000$.
\begin{figure}[!htb]
\centering
\begin{tabular}{c|c c c c c}

$\sigma$ & $\frac{1}{10}c_a $ & $ \frac{1}{2}c_a$ & $ c_a$ & $2 c_a $ & $10 c_a $\\
\hline
\\ 
$\widehat m$ & $3$ & $3$ & $3$ & $3 $ & $18$\\
  $\widehat{\mathsf G}_{n,\widehat m}$ (Hz)&$2.095\e{-3}$ &$2.095\e{-3}$ & $2.044\e{-3}$ &$2.12\e{-3}$ &$1.181\e{-1}$ \\ 
 $\widehat{\mathsf D}_{n,\widehat m}$ &$1.768\e{-4}$  &$1.784\e{-4}$ & $1.918\e{-4}$&$2.394\e{-4}$ &$3.593\e{-3}$ \\
 p-value proxy& $5\e-5$& $5\e{-5}$&$5\e{-5}$ &$5\e{-5}$ & $5.32\e{-2}$\\
\end{tabular}
\caption{{\bf Table of estimators and p-value proxys of the simulated} signals. The simulation of the null is performed with the $\ell_1$-estimate of the trend \eqref{def:l1Trend_est} of the signals.}
\label{tab:SCbis}
\end{figure}
The results are in Table~\ref{tab:SCbis}. Similarly to the first simulation, the p-value proxys for the signals with a level of noise from $\frac{1}{10}c_a$ to $2c_a$ is equal to $5\e-5$. Hence the procedure detect significant HF features where $\widehat{\mathsf G}_{n,\widehat m}\approx (2\e-3,2\e-4)$. Also for a standard deviation of the noise of $10c_a$, the p-value proxy is $5.32\e{-2}$, so that HF feature parameters are not significant enough.

\section{Empirical analysis on biological data}\label{sec:PrP}

\begin{figure}[!htb]
\centering
\includegraphics[width=0.9\textwidth]{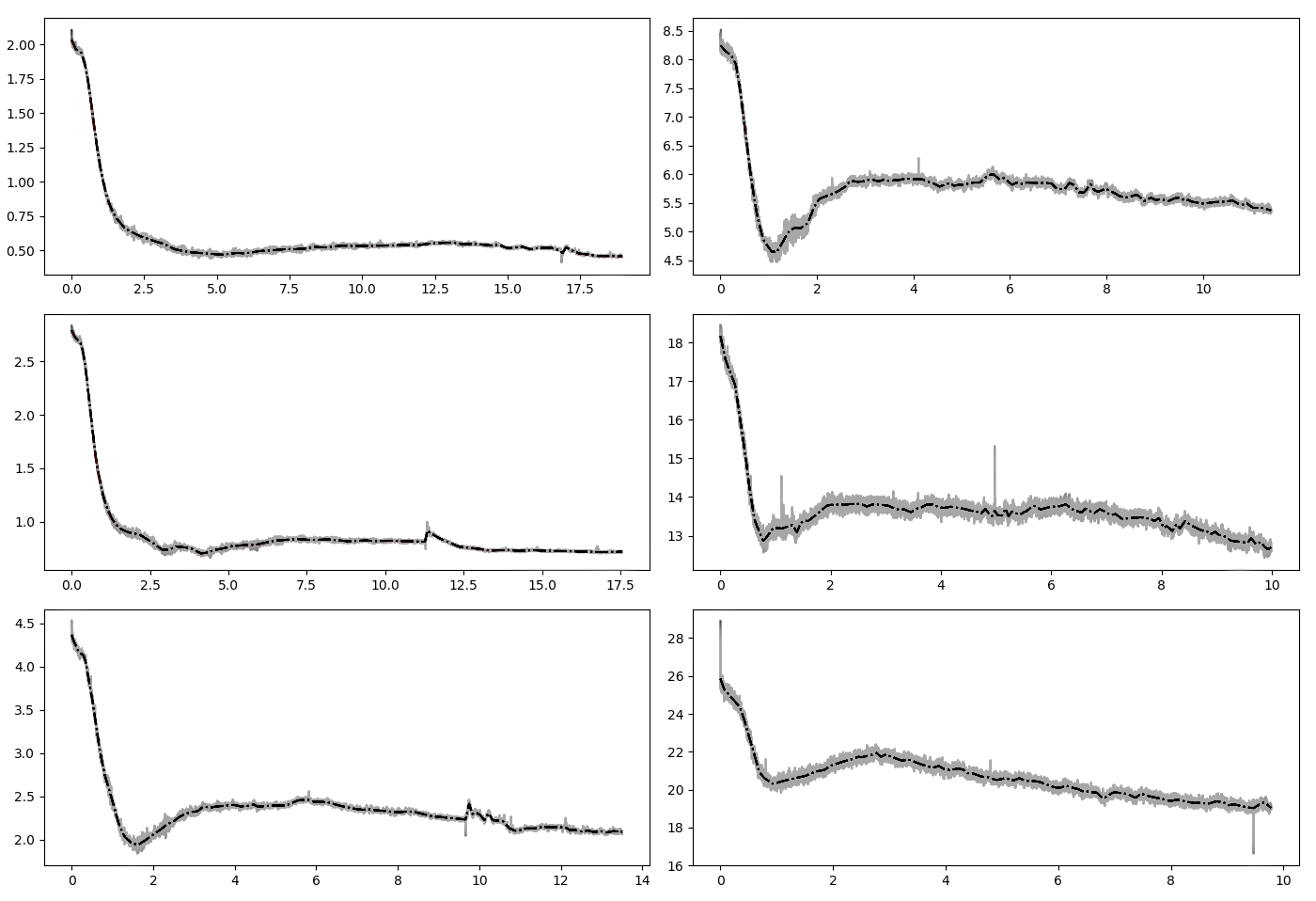}
\caption{{\footnotesize{\bf SLS experiments and trend estimates.} The x-axis is the time in hours. The parameter in the $\ell_1$-trend filtering is $\lambda=31$. (Top left) Plot of $n=32768$ samples of SLS outputs with initial concentration ($I_0$) of $0.25\mu$mol of $PrP^{Sc}$ fibrils.The dashed line is the $\ell_1$-trend estimator. (Middle left) $I_0=0.35\mu$mol  (Bottom left) $I_0=0.5\mu$mol. (Top right) $I_0=1\mu$mol. (Middle right) $I_0=2\mu$mol. (Bottom right) $I_0=3\mu$mol.  } \label{fig:RD_trend}}
\end{figure}

The Prion diseases, also known as transmissible spongiform encephalopathies (TSEs), are a group of animal and human brain diseases. The neurodegenerative processes are poorly understood and hence fatal. However the largely accepted hypothesis suggests that the infectious agent (PrPsc) is the misfolded form of the normal Prion protein (PrPc). The PrPsc forms multimeric assemblies (fibrils) which are the prerequisite for the replication and propagation of the diseases, see~\cite{prusiner1998prions}.  To follow the aggregation kinetics of these fibrils, compare it to mathematical models and get a better understanding of these diseases, several experimental  and measurement devices are used, among which the Static Light Scattering (SLS).
The Static Light Scattering (SLS) signal is an experimental measurement which describes the temporal dynamics of PrP amyloid assemblies formed in vitro, see~\cite{legname2004synthetic} - see Fig.~\ref{fig:DFMR} taken from~\cite{doumic:hal-01863748} (see Appendix~\ref{app:exper}).These signals correspond to an affine transformation of the second moment of the size distribution of protein polymers or fibrils through time, see~\cite{prigent2012efficient}:
$$\sum\limits_{i \in \mathcal{I}}i^2c_i(t) +\sigma, $$
where $\mathcal{I}$ denotes the set of the sizes of the fibrils, $c_i$ the concentration of fibrils of size $i$ which is varying with the time $t$ and $\sigma>0$ is the experimental noise ($\sigma$ can be time-dependent). At the beginning of the experiment the fibrils are  large,  containing in average several hundreds of monomers, which undergo an overall depolymerization process and leads to a decay in the signal. The experiment is carried out with six initial concentrations of fibrils (Figure~\ref{fig:RD_trend}) ranging from 0.25$\mu$mol to 3$\mu$mol; at higher initial concentrations ($0.5\mu$mol and higher), a re-polymerisation process can be observed, which may be viewed by the fact that the trend of the signal increases again before reaching a plateau. Moreover the SLS signals differ in terms of variance of noise and amplitude of oscillations (noticed by sight). We thus study  each signal independently.\\

In order to test whether the signals display HF features, we submit the observations to the statistical test described above. The denoised signal and hence the standard deviation of the noise are estimated thanks to the VisuShrink method and the median absolute deviation (cf~\cite{donoho1994ideal}, ~\cite{donoho1998minimax}) using the symmlet wavelet with 8 vanishing moments and the library Wavelab, see~\cite{buckheit1995wavelab} (the same results have been obtained with the homemade python library, see Appendix~\ref{app:LibPyth}). The trend of the signal is estimated with the $\ell_1$-trend filtering method with the parameter $\lambda=31$ ($\lambda$ is fixed qualitatively in order for the trend to include the discontinuous jumps of the SLS experiments). The results of the statistical test are summarized in Table~\ref{tab:RD}.
\begin{figure}[!htb]
\centering
\begin{tabular}{c|c c c c c c}

Concentration ($\mu$mol) & $0.25$ & $0.35$ & $0.5$ & $1$ & $2$ & $3$\\
\hline
\\
$\h{\sigma}$ & $3.553\e-3$ & $4.72\e-2$ & $1.11\e-2$ & $3.09\e-2$ & $ 8.44\e-2$ & $1.287\e-1$\\ 
$\widehat m$ & $4$ & $3$ & $5$ & $7$ & $9$ & $7$\\
 $\widehat{\mathsf G}_{n,\widehat m}$ (Hz)& $4.954\e{-3}$ &$7.53\e{-3}$ & $5.656\e{-3}$ & $8.375.\e{-3}$ & $2.698\e{-3}$ & $4.971\e{-3}$ \\ 
 $\widehat{\mathsf D}_{n,\widehat m}$ & $9.649\e{-6}$  & $1.863\e{-5}$ & $1.012\e{-4}$ & $6.526\e{-4}$ & $3.345\e{-4}$ & $1.01\e{-3}$ \\
 p-value proxy& $5\e-5$& $5\e{-5}$&$5\e{-5}$ &$5\e{-5}$ & $5\e{-5}$ & $5\e{-5}$ \\
\end{tabular}
\caption{{\bf Estimators and p-value proxys for the test of presence of HF features in the SLS experiments}}
\label{tab:RD}
\end{figure}
\begin{figure}[!htb]
\centering
\includegraphics[width=\textwidth]{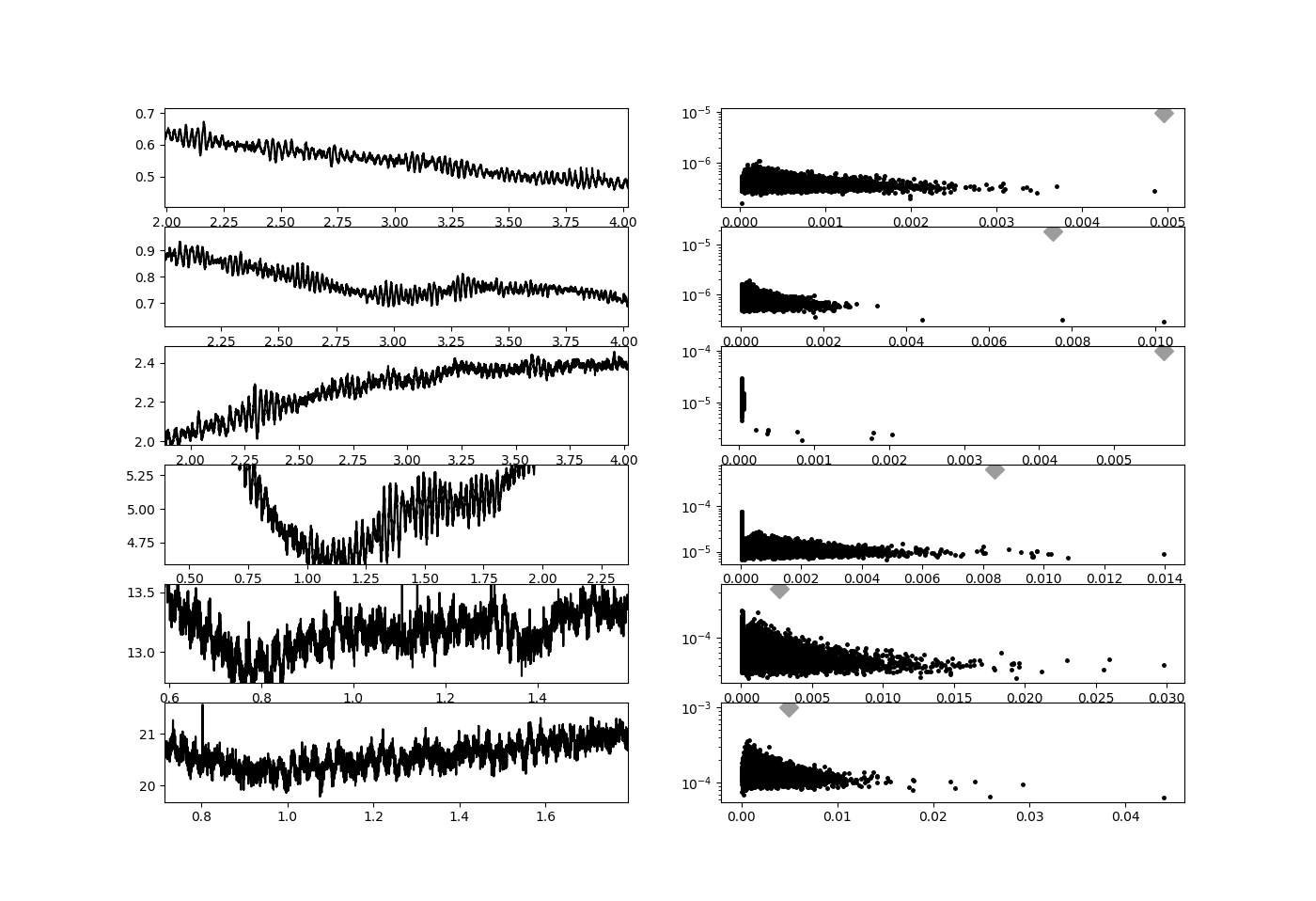}
\caption{{\footnotesize {\bf HF features of the SLS experiments and numerical results of the estimation of the HF features parameters.}(Left column) Zoom on the SLS experimentation signals with initial concentration in $\mu$mol from the top to the bottom of $I_0\in \left\lbrace0.25,\; 0.35,\; 0.5,\; 1,\; 2,\; ,3 \right\rbrace$ The x-axis is the time in Hours. (Right column) The black dots are the cloud of points $(\widehat{\mathsf G}_{n,\widehat m}^k, \widehat{\mathsf D}_{n,\widehat m}^k) $ corresponding to the simulation of the null for $k=1,\hdots,20000$. The grey diamond corresponds to the HF features parameters $(\widehat{\mathsf G}_{n,\widehat m}, \widehat{\mathsf D}_{n,\widehat m})$, defined by Definition~\ref{def: characteristics}, of the corresponding signal on the left column. The x-axis is the localization parameter of the HF features and the y-axis is the relative amplitude of the HF .
} \label{fig:RD_test}}
\end{figure}
We note that all signals display oscillations more or less pronounced (cf. Figure~\ref{fig:RD_test}). The relative amplitude of the oscillations   $\widehat{\mathsf D}_{n,\widehat m}$ differs from one signal to another for three reasons. First of all, each signal corresponds to an experiment with a specific initial concentration. The calibration of the experiments is not identical for experiments with different initial concentrations. Secondly, the signals are not on the same scale. The signal with initial concentration of $0.25\mu$mol goes from 0.5 to 2.2 in amplitude, and the signal of initial concentration of $3\mu$mol goes from 16 to 28 in amplitude. Finally, they do not have the same regularization coefficient $\widehat{m}$.
However the frequency localization parameters are comparable. In Table~\ref{tab:RD}, we note that the parameters  $\widehat{\mathsf G}_{n,\widehat m}$ are in the same range of value with a factor of less than 4 between the minimum and maximum  $\widehat{\mathsf G}_{n,\widehat m}$. Finally all the p-value proxy of the tests are equal to $5\e-5$, the tests confirm that the signals display significant HF features.\\

Through this study, we demonstrated the existence of oscillatory behavior in the SLS experiments. The immediate biochemical consequences are the coexistence of structurally distinct PrP assemblies within the same media and the unstable behavior, i.e. out of the thermo-dynamical equilibrium, of the chemical system formed by theses assemblies. Indeed the observation of oscillations in these light-scattering experiments has shed light on the existence of a complex chemical reaction network beyond the existing aggregation-fragmentation models. This has paved the way for new mechanistic models, e.g. a system of reactions which possibly involve several conformations of PrP assemblies, see~\cite{doumic:hal-01863748}, capable of explaining such phenomena. Also it has been reported that the existence of multiple conformations of PrP assemblies within an isolate contributes to the adaptation and evolution of Prion as a pathogen to a new environment and a new host, see~\cite{li2010darwinian}. Further biochemical characterizations are required to explore the dynamics of these oscillations and to establish more precise kinetic models. The methodology developed in the present work will lead to analyze and characterize with specific parameters transient oscillations. These parameters will lead to evaluate physico-chemical conditions as well as the dynamic of the present complex system.

\section{Conclusion}

In this study, we have introduced a novel method, {based on a certain nonlinear analysis of the discrete Fourier transform of a nonstationary signal}, in order to quantify high frequency features and then test whether the parameters characterizing these features may be considered as significant or not. We then tested the performance of our method on simulated and experimental data. We obtained numerical evidence of its efficiency: HF features may be detected even with a noise of comparable amplitude. Moreover, the two parameters estimated from the data that characterize the HF are informative {\it per se}: they can be used by practitioners in order to compare different experimental conditions and their influence on such transient phenomena in the signals. They may also reveal useful in the search for quantitative comparison between mechanistic models, such as the one proposed in~\cite{doumic:hal-01863748}, and experimental data. Our testing procedure for detecting HF feature is based on the projection of the signal in a discrete Fourier basis. A further step, in order to localize them, would be to define them in terms of  wavelet bases \cite{donoho1994threshold,donoho1998minimax,donoho1995wavelet,donoho1994ideal}. The number of parameters will then be equal to three (one for the resolution, one for the amplitude and one for the localisation on the time-scale), and the  test of hypothesis has to be extended to this framework. This is a direction for future work.\\


\begin{appendix}
\section{Materials and methods of the depolymerisation experiment} 
\label{app:exper}

Formation of amyloid fibrils: PrP amyloid fibrils were formed using the manual setup protocol described previously in~\cite{breydo2008methods}. Fibril formation was monitored using a ThT binding assay, see~\cite{breydo2008methods}. Samples were dialysed in 10 mM sodium acetate, pH 5.0. Then fibrils were collected by ultracentrifugation and resuspended in 10 mM sodium acetate, pH 5.0. A washing step was performed by repeating the ultracentrifugation and resuspension steps in 10 mM sodium acetate, pH 5.0.
Static light scattering: Static light scattering kinetic experiments were performed with a thermostatic homemade device using a 407-nm laser beam. Light-scattered signals were recorded at a 112$^0$ angle. Signals were processed with a homemade MatLab program. All experiments have been performed at 55$^0$C in a 2mmX10mm cuve. 

\section{Library in python to implement the numerical simulation}\label{app:LibPyth}
The numerical simulations have been made with the Python library accessible at \textit{https://github.com/mmezache/HFFTest}. The functions of the library are explicitely commented in the file "README.md". The functions are organized in four categories in the library:
\begin{enumerate}
\item the procedure to compute the HF features parameters,
\item the procedure to simulate the null hypothesis,
\item the Monte Carlo procedure to compute the p-value proxy,
\item the procedure to compute test signals such as the ones displayed in Figures~\ref{fig:DFT_lowFreqOsc},  \ref{fig:DFT_lowFreqjump}, \ref{fig:SC_Signal}.
\end{enumerate}
The file "ExampleHFF.py" is a python program which computes the complete procedure for a test signal. The users may change at will the following parameters:
\begin{itemize}
\item the length of the signal,
\item the standard deviation of the noise,
\item the amplitude of the oscillations,
\item the parameter of the $\ell^1$-trend filtering,
\item the number of iteration of the Monte Carlo procedure,
\item the choice of the test signal.
\end{itemize}
The program displays the test signal obtained, the trend estimate, the cloud of points corresponding to the HF features of the null (blue dots) and the point corresponding to the HF features of the tested signal (grey dot), and the single-sided amplitude spectrum of the signal which emphasizes the points where the computations of the HF features are performed (cf Figure~\ref{fig:scheme_par_osc}).\par
The computational time may be significantly long if the number of iterations of the Monte Carlo procedure is large (over 100). However the Monte Carlo procedure can be computed in a parallelized framework which reduces drastically the computational time.\par
Moreover the automatic choice of the smoothing parameter $\widehat{m}$ is efficient for signals which display oscillations of "high" frequency, i.e. if the spike corresponding to the oscillations in the single-sided amplitude spectrum is located away from the low-frequency components (cf Section~\ref{sec:num} and Example~2 in "ExampleHFF.py"). The procedure was designed to identify oscillations "hidden" in the noise, a situation which corresponds to the experimental signals. If the signal tested has oscillations located in the low frequencies, the users are advised to fix the smoothing parameters (see Example~1 in "ExampleHFF.py"). 

\section{Trend estimation}
\label{app:trend}
We detail here the reasons for the choice of $\ell^1$ trend filtering method to estimate the trend of the signal.

Trend estimation or filtering for mimicking a signal with HF features removed has many applications and hence it has been extensively studied. It has given rise to the smoothing and filtering methods such as the moving average~\cite{xu2005quantifying}, smoothing splines~\cite{reinsch1967smoothing}, Hodrick-Prescott filtering~\cite{ravn2002adjusting}, $\ell_1$-trend filtering~\cite{kim2009ell_1} and so on. The trend is considered as the general shape of a signal or a time series. Although  the trend is often understood and perceived intuitively, its estimator relies on the definitions given to the trend. The differences between the various definitions of the trend are a matter of interpretation. Considering the different definitions of the trend, the choice of the method to estimate this component is more likely qualitative. In the following, the trend is considered as the underlying slowly varying component of the signal and we choose the $\ell_1$-trend filtering method described in ~\cite{kim2009ell_1} to estimate it. The estimator of $\widehat x_{\lambda, n}^{(0)}$ as a $n$-dimensional vector is then the solution of the following optimisation problem (Equation~\eqref{def:l1Trend_est} above):
\begin{equation*}
\widehat x_{\lambda, n}^{(0)} \in \argmin\limits_{x\in\mathbb{R}^n}\frac{1}{2}\sum\limits_{i=1}^{n-1} (y_i^n-x_i^n)^2+\lambda\sum\limits_{i=1}^{n-2}|x_{i-1}^n-2x_i^n+x_{i+1}^n|,
\end{equation*}
where $\lambda\geq 0$ is a regularisation parameter which controls the trade-off between the smoothness of $\widehat x_{\lambda, n}^{(0)}$ and the residual $\sum_{i = 0}^{n-1}\big(\widehat x_{\lambda, n,i}^{(0)}-y_i^n\big)^2$. We note that the second term $ \sum\limits_{i=1}^{n-2}|x_{i-1}^n-2x_i^n+x_{i+1}^n|$ is the $\ell^1 $-norm of the second order variations of the sequence $(x^n)$ (i.e. the discretization of the corresponding $L^1$-norm of the second derivative of a function). Moreover, for any sequence $(x^n)$, we have $|x_{i-1}^n-2x_i^n+x_{i+1}^n|=0$ for every $i$ if and only if $x_i^n=\alpha i+\beta$ for two real parameters $\alpha$ and $\beta$. Thus only an affine function has its $\ell^1$-norm equal to 0. Hence this method gives an estimator of the trend such that:
\begin{enumerate}
\item[(i)] $\widehat x_{\lambda, n}^{(0)}$ is computed numerically in $\mathcal O(n)$ operations,
\item[(ii)] as $\lambda \rightarrow 0$, $\max_{0 \leq i \leq n-1}|\widehat x_{\lambda, n}^{(0)}-y_i^n| \rightarrow 0$, the estimator converges to the original data,
\item[(iii)] as $\lambda \rightarrow \infty$, the estimator converges to the best affine fit of the observations. This convergence happens for a finite value of $\lambda$. 
\item[(iv)] $\widehat x_{\lambda, n}^{(0)}$ is piecewise linear, i.e. there are indices $ 0=j_1 <j_2<\hdots<j_K=n-1 $ for which:
$$\widehat x_{\lambda, n,i}^{(0)}=\alpha_k i +\beta_k,\quad j_k<i<j_{k+1},\quad k=1,\hdots,K-1. $$
\end{enumerate} 

\begin{figure}[!htb]
\centering
\includegraphics[height=0.5\linewidth]{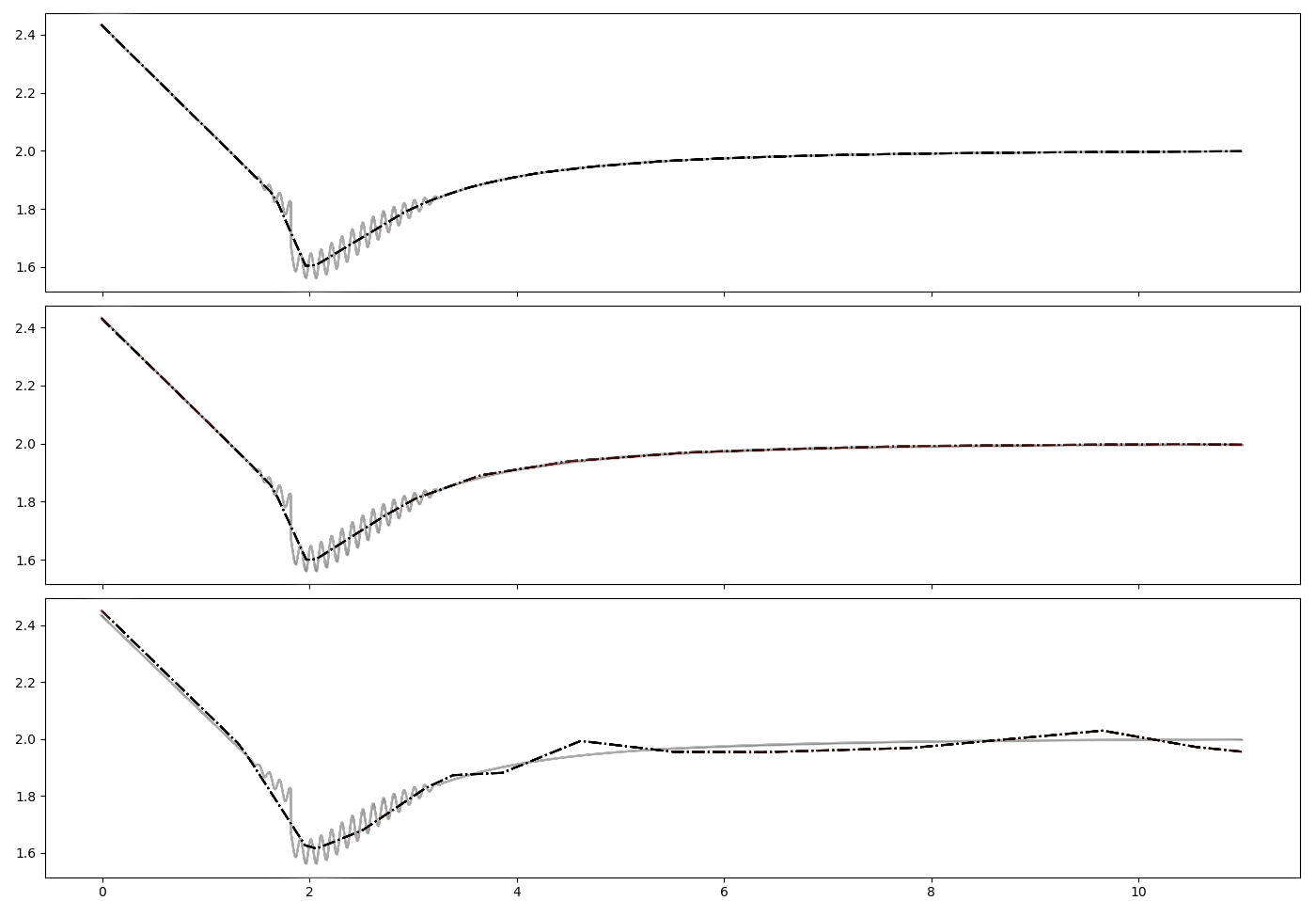}
\caption{{\footnotesize{\bf Numerical estimation of the trend on the test signals.} The x-axis is the time in hours.The parameter in the $\ell_1$-trend filtering is $\lambda=301$. (Up) Plot of $(P_i+O_i)_{0\leq i \leq 10^5}$ in \eqref{Def:SignalSC} with parameters $c_1=0.4,\; c_2=c_3=c_4=2,\; \frac{p}{2}=q=3, t_0=1.43,\; t_1=3.30$ $c_a=0.05,\; c_f=10$. The dashed line is the $\ell_1$-trend estimator when $\sigma=\frac{1}{10}c_a$. (Middle) The dashed line is the $\ell_1$-trend estimator when $\sigma=c_a$. (Down )The dashed line is the $\ell_1$-trend estimator when $\sigma=10c_a$. } \label{fig:SC_trend}}
\end{figure}

The $\ell_1$-trend filtering method is well suited to extract the trend components of the signals studied in Section~\ref{sec:num}. Since the signals display singularities such as discontinuous jumps, the trend extracted is well approximated by a piecewise linear function. Moreover the HF features in the signals are components looking like sine waves and varying at an intermediate pace. However interpolating a sine wave by a piecewise linear function requires a fine scale and thus the parameter $\lambda$ has to be close to $0$. Rising slightly the value of $\lambda$ allows us to capture the trend without the HF features. Moreover there exists a nonnegative threshold $\lambda_{\max}$ such that $\widehat x_{\lambda_{\max}, n}^{(0)}$ is the trend estimator corresponding to the best affine fit, see \cite{kim2009ell_1}. It implies that the choice of $\lambda$ is restricted to the bounded open interval $(0, \; \lambda_{\max}).$ Since there is no optimal criterium to choose $\lambda$, the choice of the parameter is qualitative and motivated empirically (see Section~\ref{sec:num}). As displayed in Figure~\ref{fig:SC_trend}, when noise increases, the trend is less robustly estimated.
\section{Noise estimation}\label{app:noise}
In this appendix, we detail the method and the parameters we chose to estimate the noise of the signal.
Noise estimation is linked to signal denoising and has been extensively studied. The method chosen here  is the median absolute deviation and the denoised signal is obtained thanks to the wavelet shrinkage methods, see {\it e.g.}~\cite{donoho1994threshold,donoho1998minimax,donoho1995wavelet,donoho1994ideal}. 

We assume that our data $y=(y_i)_{0\leq i\leq n-1}$ are such that $n=2^{J+1}$ for $J>0$. We then consider an orthogonal wavelet transform matrix $\mathcal{W}$ for a given filter. Choosing wavelets (e.g. Coiflet, Daubechies, Haar) and varying the combinations of parameters M (number of vanishing moments), S (support width) and $j_0$ (low-resolution cut-off) one may construct various orthogonal matrices $\mathcal{W}$ (see for details~\cite{mallat1999wavelet}, chapter 7). In this paper we use the Symmlet with parameter $8$ which has $M=7$ vanishing moments and support length $S=15$.
The wavelet coefficients of $y$ are denoted by $w$ and
$$ w=\mathcal{W}x+\sigma \widetilde \xi,$$
where $\widetilde \xi = \mathcal{W}\xi$ is a standard Gaussian random vector by orthogonality of $\mathcal W$. 
For convenience, we index dyadically the vector of the wavelet coefficients
$$w_{j,k}\quad j=0,\hdots, J,\quad k=0,\hdots,2^j-1. $$
We make the legitimate assumption that empirical wavelet coefficients at the finest resolution level $J$ are essentially pure noise. Hence the standard deviation estimator $\widehat \sigma_n$ is the median absolute deviation
\begin{equation}
\widehat \sigma_n=\frac{\text{median}(w_{J,\cdot})}{\Phi^{-1}(3/4)},
\label{def:sigma_MAD}
\end{equation}
\noindent
where $\Phi^{-1}(\cdot)$ is the inverse of the cumulative distribution function for the standard normal distribution. Thus $\widehat \sigma_n$ is a consistent estimator of $\sigma$. It is interesting to note that further computations give the VisuShrink estimator $\widehat x_n$ of the signal $(x_i^n)_{0 \leq i \leq n-1}$
\begin{equation}
\widehat x_n=\mathcal{W}^\top\widehat{w}^{n,j_0}\mathcal{W},
\label{def:f_Visu}
\end{equation}
\noindent
where $j_0$ denotes a low resolution cut-off and $\widehat{w}^{n,j_0} $ is the estimator in the wavelet domain
\begin{equation*}
\widehat{w}^{n,j_0} =\left\{\begin{array}{l@{}l}w_{j,\cdot}&\quad j<j_0\\
\text{sign}(w_{j,\cdot})\left(|w_{j,\cdot}|-\widehat \sigma_n(2\log n)^{1/2}\right)_+&\quad j_0\leq j\leq J
\end{array} \right. .
\end{equation*}
\par
The first reason that motivates this choice is that shrinkage methods attempt to remove whatever noise is present and retain whatever signal is present regardless of the frequency, see~\cite{donoho1994ideal}. The goal of this paper is to estimate HF features in noisy signals. However other methods of noise removal such as low-pass filters are based on frequency-dependent estimators, and this can also impact and distort the results of the HF feature procedure. The second reason is that these methods are data-driven and no specific assumptions on the signal are required. Wavelet shrinkage is spatially adapted and the method is efficient for a wide variety of signals even when the signals exhibit spatial inhomogeneities. Moreover, these methods are proven to be nearly optimal for the mean squared error criterion when the smoothness of the original signal is unknown, see~\cite{donoho1995wavelet}.

\section{Definition of the test signal}\label{app:test}

For the general trend, we choose the Lennard Jones potential, see~\cite{jones1924determination}, since we notice that its DFT is not monotonously decreasing in the low frequency range (see Figure~\ref{fig:DFT_lowFreqOsc}) and that it displays a similar shape as  the experimental signals presented in Section~\ref{sec:PrP}. The Lennard Jones potential is defined by $P_i$:
$$P_i=\bigg(c_1\left[\left(\frac{c_2}{i}\right)^{p}-c_3\left(\frac{c_2}{i}\right)^{q} \right] + c_4\bigg). $$
 Since this potential is not defined at $0$, we link the potential to an affine function. Hence we introduce the index $j$ ($0<j<n-1$) which connects the potential to the affine function. We denote the trend by the vector $(T_i)_{0\leq i\leq n-1}$:
$$T_i=\left(\frac{P_{j+1}-P_j}{j+1} i + P_j\right) \mathds{1}_{\lbrace0\leq i\leq j\rbrace} +P_i\mathds{1}_{\lbrace j+1\leq i \leq n-1\rbrace}.$$ 

The HF features in the test signal correspond to sine waves and are located at a specific time interval. Hence we introduce the indices $0<j_0<j_1<n-1$ which localize the oscillations in the signal, and we define the oscillations by the vector $ (O_i)_{0\leq i\leq n-1}$:
\begin{equation}
O_i=c_{a}(i-j_0)(j_1-i)\sin(2\pi c_{f}i)\left(\frac{4}{(j_1-j_0)^2}\right)\mathds{1}_{\lbrace j_0\leq i\leq j_1  \rbrace}\label{def:Osc_SignalSC}
\end{equation}
\noindent
where $c_a$ (resp. $c_f$) is the parameter for the amplitude ( resp. the frequency) of the oscillations. 
\end{appendix}

\bibliographystyle{ieeetr}
\nocite{*}
\bibliography{OscillationsDetection}

\end{document}